\documentclass{emulateapj}
\usepackage{apjfonts}

\newcommand{\mscl}{$\rm MS\,1054-03$}
\newcommand{\etal}{et al.~}

\newcommand{\masssol}{$\rm ~M_{\odot}$}

\newcommand{\tauint}{$\rm \tau_{300\,Myr}$}

\slugcomment{Accepted for publication in Astrophysical Journal}

\shorttitle{Old, dusty, and massive galaxies at $z > 2$}
\shortauthors{F\"orster Schreiber \etal}

\begin{document}

\title{A Substantial Population of Red Galaxies at $z > 2$: \\
       Modeling of the Spectral Energy Distributions of an Extended Sample
       \altaffilmark{1}}

\author{N. M. F\"orster Schreiber\altaffilmark{2,3},
	P. G. van Dokkum\altaffilmark{4},
        M. Franx\altaffilmark{2},
	I. Labb\'e\altaffilmark{2},
	G. Rudnick\altaffilmark{5},
	E. Daddi\altaffilmark{6},
        G. D. Illingworth\altaffilmark{7},
	M. Kriek\altaffilmark{2},
	A. F. M. Moorwood\altaffilmark{6},
	H.-W. Rix\altaffilmark{8},
	H. R\"ottgering\altaffilmark{2},
        I. Trujillo\altaffilmark{8},
	P. van der Werf\altaffilmark{2},
	L. van Starkenburg\altaffilmark{2},
	S. Wuyts\altaffilmark{2}}
%\email{forster@strw.leidenuniv.nl}

\altaffiltext{1}{Based on observations collected at the European
   Southern Observatory, Paranal, Chile (ESO LP Programme 164.O-0612).
   Based on observations with the NASA/ESA Hubble Space Telescope
   obtained at the Space Telescope Science Institute, which is operated by
   the AURA, Inc., under  NASA contract NAS5-26555.
   Based on observations obtained at the W. M. Keck Observatory,
   which is operated jointly by the California Institute of Technology and
   the University of California.}
\altaffiltext{2}{Leiden Observatory, PO Box 9513, 2300 RA Leiden,
                 The Netherlands}
\altaffiltext{3}{Present address: 
                 Max-Planck-Institut f\"ur extraterrestrische Physik,
                 Giessenbachstrasse, D-85748 Garching, Germany}
\altaffiltext{4}{Department of Astronomy, Yale University, P.O. Box 208101, 
                 New Haven, CT 06520-8101}
\altaffiltext{5}{Max-Planck-Institut f\"ur Astrophysik, Postfach 1317,
                 D-85741 Garching, Germany}
\altaffiltext{6}{European Southern Observatory, Karl-Schwarzschildstr. 2,
                 D-85748 Garching, Germany}
\altaffiltext{7}{UCO/Lick Observatory, University of California,
                 Santa Cruz, CA 95064}
\altaffiltext{8}{Max-Planck-Institut f\"ur Astronomie, K\"onigstuhl 17,
                 D-69117 Heidelberg, Germany}

\begin{abstract}

We investigate the nature of the substantial population of high-redshift
galaxies with $J_{\rm s} - K_{\rm s} \geq 2.3$ colours recently discovered
as part of our Faint InfraRed Extragalactic Survey (FIRES).
This colour cut efficiently isolates galaxies at $z > 2$ with red
rest-frame optical colors (``Distant Red Galaxies,'' or ``DRGs'').
We select $J_{\rm s} - K_{\rm s} \geq 2.3$ objects in both FIRES fields, the
$\approx 2.5^{\prime} \times 2.5^{\prime}$ Hubble Deep Field South (HDF-S) and
the $\approx 5^{\prime} \times 5^{\prime}$ field around the \mscl\ cluster
at $z = 0.83$;
the surface densities at $K_{\rm s, Vega} < 21~{\rm mag}$ are
$\rm 1.6 \pm 0.6$ and $\rm 1.0 \pm 0.2~arcmin^{-2}$, respectively.
We here discuss a sub-sample of 34 DRGs at $2 \leq z \leq 3.5$:
11 at $K_{\rm s, Vega} < 22.5~{\rm mag}$ in HDF-S and
23 at $K_{\rm s, Vega} < 21.7~{\rm mag}$ in the \mscl\ field.
This sample enables for the first time a robust assessment of the
population properties of DRGs.
We analyze the $\rm \lambda = 0.3 - 2.2~\mu m$ spectral energy distributions
(SEDs) constructed from our very deep near-infrared (NIR) and optical imaging
collected at the ESO Very Large Telescope and from the Hubble Space Telescope.
We develop diagnostics involving the $I_{814} - J_{\rm s}$, $J_{\rm s} - H$,
and $H - K_{\rm s}$ colours to argue that the red NIR colours of our DRG
sample cannot be attributed solely to interstellar dust extinction and
require for many the presence of an evolved stellar population with a
prominent Balmer/4000\,\AA\ break.
In the rest-frame, the optical colours of DRGs fall within the envelope of
normal nearby galaxies and the ultraviolet colours suggest a wide range in
star formation activity and/or extinction.  This is in stark contrast with
the much bluer and more uniform SEDs of Lyman-break galaxies (LBGs).
From evolutionary synthesis modeling assuming constant star formation (CSF),
we derive for the DRGs old ages, large extinctions, and high stellar masses,
mass-to-light ratios, and star formation rates.
For solar metallicity, a Salpeter IMF between 0.1 and 100\masssol,
and the Calzetti et al. extinction law, the median values for the
HDF-S (\mscl\ field) sample are 1.7 (2.0)~Gyr, $A_{V} = 2.7$ (2.4)~mag,
$M_{\star} = 0.8$ (1.6)~$\times 10^{11}~{\rm M_{\odot}}$,
$M_{\star}/L_{V,\star} = 1.2$ (2.3)~${\rm M_{\odot}\,L_{V,\odot}^{-1}}$,
and ${\rm SFR} = 120$ (170)~${\rm M_{\odot}\,yr^{-1}}$.
Models assuming exponentially declining SFRs with $e$-folding timescales in
the range $\rm 10~Myr - 1~Gyr$ generally imply younger ages, lower extinction,
and lower SFRs, but similar stellar masses within a factor of two.
Compared to LBGs at similar redshifts and rest-frame $V$-band luminosities,
DRGs are older, more massive, and more obscured for any given star
formation history.
For the entire sample of $K_{\rm s}$-band selected galaxies in the
FIRES fields at $2 \leq z \leq 3.5$ and to the above magnitude limits,
we find that the derived ages, extinctions, and stellar masses increase
with redder $J_{\rm s} - K_{\rm s}$ colours.
Although the rest-frame optical colours of the DRGs are similar to those
of local normal galaxies, the derived properties are quite different;
detailed studies of this new $z > 2$ population may significantly enhance
our understanding of how massive galaxies assembled their stellar mass.
\end{abstract}


\section{INTRODUCTION}   \label{Intro}

Understanding how galaxies formed and evolved is a central challenge of modern
astronomy.  In the past decade, spectacular progress in instrumentation and
observing techniques has enabled major advances as the high-redshift universe
has been opened up for research.
Our current view has been largely influenced by the discovery of an abundant
population of actively star-forming galaxies at redshifts $z \sim 3 - 4$
selected by the efficient Lyman-break technique \citep{Ste93, Ste96a, Ste96b}.
These Lyman-break galaxies (LBGs) are among the best-studied classes of
high-redshift objects: large samples have been spectroscopically confirmed
at $z \sim 3 - 4$ \citep[e.g.][]{Ste99} and extensive investigations have
focussed notably on their stellar populations, star formation histories,
chemical abundances, kinematics, and clustering properties
\citep*[e.g.][]{Gia98, Pet01, Pap01, Sha01, Sha03, Erb03}.
LBGs dominate the ultraviolet (UV) luminosity density at high redshift and
their number density is comparable to that of $L^{\star}$ galaxies locally,
making them a major constituent of the early universe.  In the context of the
widely favoured hierarchical scenarios of galaxy formation, they are thought
to be the progenitors of present-day massive galaxies in groups and clusters
\citep[e.g.][]{Bau98}.

Yet, by construction the Lyman-break technique relies on a strong Lyman
discontinuity in the rest-frame far-UV and is necessarily biased towards
relatively unobscured galaxies with intense recent or on-going star
formation activity.  The typical stellar ages derived for LBGs are of
a few $\rm 10^{8}~yr$ with moderate extinction $A_{V} < 1~{\rm mag}$
\citep{Pap01, Sha01}.  Older and more quiescent systems at $z \sim 3$ that have
formed the bulk of their stars at $z \ga 4$ or, alternatively, more obscured
galaxies may have escaped detection in optical surveys \citep*[e.g.][]{Fer02}.
With the advent of $\rm 8-10~m$-class telescopes and the development of
sensitive near-infrared (NIR) instruments equipped with large-format detectors,
it has become possible to extend very deep surveys to longer wavelengths and
access the rest-frame optical emission of sources at $z \sim 1 - 4$.
Compared to the rest-frame UV, the rest-frame optical spectral energy
distribution (SED) of galaxies is less affected by the light of young
massive stars and by dust extinction, and better traces longer-lived
stars that dominate the stellar mass.

Colour criteria involving NIR bandpasses can now be applied to identify
new populations at $z > 1$.  The long-standing debate over the formation
of massive early-type galaxies has driven considerable interest in the
discovery of, and subsequent systematic searches for, red objects such
as the ``extremely red objects'' (EROs) generally defined by
$R - K > 5$ or $I - K > 4~{\rm mag}$
\citep*[e.g.][among others]
{Els88, Hu94, Tho99, Yan00, Sco00, Dad00, McC01, Smi02}.
In our own Faint InfraRed Extragalactic Survey \citep[FIRES;][]{Fra00},
based on very deep NIR imaging of the Hubble Deep Field South (HDF-S)
and of the field around the $z = 0.83$ cluster \mscl, we have identified
significant numbers of fairly bright ($K_{\rm s} \approx 19 - 22.5~{\rm mag}$)
candidate $z > 2$ galaxies selected from their $J_{\rm s} - K_{\rm s} \geq 2.3$
colours.  Analysis of the HDF-S sample suggests that this population makes
a comparable contribution to the stellar mass density at $z \sim 3$ as LBGs
\citep{Rud03, Fra03} and hence may be a substantial component in terms of
stellar mass.  Remarkably, there are far fewer such bright red objects per
unit area in the HDF-North (e.g. \citealt{Dic00}; see also \citealt{Lab03}).
Cosmic variance may however largely explain these differences since both 
Hubble Deep Fields are small and the $J_{\rm s} - K_{\rm s} \geq 2.3$
population may be strongly clustered \citep{Dad03}.  Candidate high-redshift
galaxies with unusually red $J_{\rm s} - K_{\rm s}$ colours have been reported
by other authors as well \citep[e.g.][]{Sco00, Hal01, Tot01, Sar01, Sar03}.
However, the focus has been on those objects with the most extreme
colours, which are rarer and mostly at the faintest magnitudes.

An immediate question is what causes the red colours of these objects.
Very red NIR colours can be produced by an evolved population at $z \geq 2$
due to the stellar photospheric Balmer/4000\,\AA\ break redshifted into the $J$
band and beyond or by high levels of extinction in galaxies possibly at lower
redshifts.  The presence of a highly obscured active galactic nucleus (AGN)
is another possibility \citep[e.g.,][]{Nor02, Koe04}.  While it appears that a
combination of stellar population aging and extinction effects is required to
explain the observed colours \citep[e.g.][]{Dic00, Tot01, Hal01, Fra03, Sar03},
the results so far remain inconclusive due to the scarcity of systematic
studies of large samples and to insufficient observational constraints.  

In this paper, we take advantage of our FIRES data set to address the above
issue from the $\rm \lambda = 0.3 - 2.2~\mu m$ broad-band SEDs of 34 objects
with $J_{\rm s} - K_{\rm s} \geq 2.3$ selected from both fields surveyed.
The addition of the \mscl\ objects triples the original HDF-S sample.  We
focus on the ensemble properties of the $J_{\rm s} - K_{\rm s}$ red galaxies.
\citet{Dok03, Dok04} present results of our follow-up optical and NIR
spectroscopy of a bright sub-sample in the \mscl\ field, which are
particularly relevant to this work.

We briefly describe the data in \S~\ref{Sect-data}.
We discuss the selection criteria applied to construct the
$J_{\rm s} - K_{\rm s}$ selected samples in \S~\ref{Sect-sample}.
In \S~\ref{Sect-prop}, we analyze the properties of the SEDs
to investigate the stellar populations and dust obscuration.
We model the SEDs using evolutionary synthesis in \S~\ref{Sect-models}
to constrain quantitatively the stellar ages and masses, the extinction,
and the star formation rates.
In \S~\ref{Sect-discussion}, we interpret our results and compare the
derived properties to those of LBGs and of $K$-band-selected objects at
similar redshifts in the FIRES fields.
We summarize the paper and main implications in \S~\ref{Sect-conclu}.
For convenience, we adopt the term ``distant red galaxies,'' or ``DRGs,''
introduced by \citet{Dok04} to designate candidate $z > 2$ galaxies with
$J_{\rm s} - K_{\rm s} \geq 2.3$ colours.
All magnitudes are expressed in the Vega-based photometric system except
when ``AB'' indicates reference to the AB system \citep{Oke71}.  Throughout,
we assume a $\Lambda$-dominated cosmology with $\Omega_{\rm m} = 0.3$,
$\Omega_{\Lambda} = 0.7$, and $H_{0} = 70~h_{70}~{\rm km\,s^{-1}\,Mpc^{-1}}$.

\section{DATA}       \label{Sect-data}

The data were obtained as part of FIRES \citep{Fra00}, a public NIR
survey of the HDF-S and \mscl\ fields carried out at the European
Southern Observatory (ESO) Very Large Telescope (VLT).
The observations, data reduction, and source catalogues are described in
detail by \citet{Lab03} for HDF-S and N.M. F\"orster Schreiber \etal
(2004, in preparation) for the \mscl\ field
\,\footnote{The reduced images, photometric catalogues, and photometric
redshifts are available online through the FIRES homepage at
{\tt http://www.strw.leidenuniv.nl/\~{ }fires}.}.

Briefly, the NIR observations were taken in the $J_{\rm s}$, $H$, and
$K_{\rm s}$ bands with the VLT Infrared Spectrograph And Array Camera
\citep[ISAAC;][]{Moo98}.
For HDF-S, a total of 103~h integration time was spent in a single
$\approx 2.5^{\prime} \times 2.5^{\prime}$ pointing covering the Hubble
Space Telescope (HST) WFPC2 main field.  We complemented the NIR data with
the publicly available deep optical WFPC2 imaging in the $U_{300}$, $B_{450}$,
$V_{606}$, and $I_{814}$ bands \citep{Cas00}.  For the \mscl\ field, 77~h of
ISAAC integration time was obtained in a $\approx 5^{\prime} \times 5^{\prime}$ 
mosaic of four pointings.  At optical wavelengths, we used existing HST WFPC2 
mosaics in the $V_{606}$ and $I_{814}$ bands \citep{Dok00} and collected
additional imaging in the Bessel $U$, $B$, and $V$ bands with the VLT FORS1
instrument.  With HDF-S, FIRES achieves the currently deepest ground-based
NIR imaging and the deepest $K$-band map to date, even from space.  With
the \mscl\ field, the area surveyed is nearly five times wider, down to
$\approx 0.7~{\rm mag}$ brighter magnitudes.

The sources were detected in the $K_{\rm s}$ band using version 2.2.2 of the
SExtractor software \citep{Ber96}.  For consistent photometry across all bands,
the fluxes were measured on the maps convolved to a common spatial resolution
of full-width at half maximum FWHM = 0.48\arcsec\ for HDF-S and 0.69\arcsec\
for the \mscl\ field, matching to the map of poorest seeing in each data set
($H$ and $U$ band, respectively).
From the curve-of-growth analysis of the average stellar profile in each
band, the fractional enclosed flux agrees to within 3\% at diameters
$0.7^{\prime\prime} \leq d \leq 2.0^{\prime\prime}$ for HDF-S and 2\% at
$1.0^{\prime\prime} \leq d \leq 2.0^{\prime\prime}$ for the \mscl\ field,
where the diameters range from the smallest ($\rm 1.5 \times FWHM$ of the
matched PSFs) to the largest apertures relevant for the colour measurements.
Colours and SEDs used in this work are based
on measurements in custom isophotal apertures defined from the detection map,
hereafter referred to as ``colour'' fluxes or magnitudes.  Total magnitudes
in the $K_{\rm s}$ band were computed in apertures based on auto-scaling
apertures \citep{Kro80} for isolated sources and adapted isophotal apertures
for blended sources.  The photometric uncertainties were derived empirically
from simulations on the maps.  The total $5\sigma$ limiting magnitudes for
point sources are $K_{\rm s}^{\rm tot} = 23.8~{\rm mag}$ for HDF-S and
23.1~mag for the \mscl\ field.  We will refer to the objects by their
number in catalogue versions v3.0e for HDF-S and v3.1b for the \mscl\ field,
prefixed with ``$\rm HDFS-$'' and ``$\rm MS-$.''

Wherever appropriate, we corrected for the Galactic extinction along the
HDF-S and \mscl\ lines of sight \citep*[$A_{V} = 0.09$ and 0.11~mag;][]{Sch98}.
We also accounted for the gravitational lensing by the foreground \mscl\
cluster based on the lensing model constructed by \citet*{Hoe00} from weak
lensing analysis.  The average magnification factor for all $K_{\rm s}$-band
selected objects considered in this paper is 1.18 (or 0.18~mag).  We emphasize
that the colours and SED shapes are unaffected since lensing is achromatic.

Photometric redshifts $z_{\rm ph}$ were determined by applying an algorithm
involving linear combinations of redshifted spectral templates of galaxies of
various types, as described by \citet{Rud01, Rud03}.  Monte-Carlo simulations
were used to estimate the $z_{\rm ph}$ errors accounting for uncertainties
in the fluxes as well as template mismatch, and reflecting not only the
confidence interval around the best solution but also the presence of
secondary solutions.  Comparison with available spectroscopic redshifts
$z_{\rm sp}$ imply an accuracy of $\delta z \equiv
\langle |z_{\rm sp} - z_{\rm ph}| / (1 + z_{\rm sp}) \rangle = 0.074$
for both fields.  For the $z \geq 2$ regime of interest in this work
(see \S~\ref{Sub-sample}), $\delta z = 0.051$ based on the 20
$K_{\rm s}$-band selected sources with $z_{\rm sp}$ determinations.

\section{NEAR-INFRARED SELECTED RED $z \geq 2$ SAMPLES} 
         \label{Sect-sample}

\subsection{The $J_{\rm s} - K_{\rm s}$ Colour Criterion}
            \label{Sub-criterion}

To identify red candidate high-redshift galaxies, or DRGs,
we applied the criterion $J_{\rm s} - K_{\rm s} \geq 2.3~{\rm mag}$
discussed by \citet{Fra03}.
This criterion was specifically designed to target the stellar photospheric
Balmer and 4000\,\AA\ breaks at redshifts above 2, which are the most
prominent features in the rest-frame optical continuum SED of galaxies.
The Balmer discontinuity at 3650\,\AA\ is strongest in A-type stars and,
more generally, is indicative of stellar ages $\rm \sim 10^{8} - 10^{9}~yr$.
At older ages, the 4000\,\AA\ break characteristic of cooler stars with
types later than about G0, and strongest in giants and supergiants, becomes
dominant.  It is due to the sudden onset of metallic and molecular opacity
bluewards of 4000\,\AA\ produced notably by \ion{Ca}{2} H$+$K, \ion{Fe}{1},
\ion{Mg}{1}, and CN lines.

Based on synthetic spectral templates computed with the \citet{Bru03} models
(see \S~\ref{Sect-models}), unextincted single-age stellar populations
older than $\rm \approx 250~Myr$ or passively evolving galaxies formed at
$z_{\rm f} \ga 5$ satisfy the criterion at $2 \la z \la 4$, where the
Balmer/4000\,\AA\ break is redshifted into the NIR regime.
Galaxies forming stars at a constant rate have a $J_{\rm s} - K_{\rm s}$
colour redder than 2.3~mag only if they are $\rm \ga 1~Gyr$ old and
moderately obscured at $z \ga 2$, or younger and highly obscured with
$A_{V} > 2~{\rm mag}$ at $z \ga 1$.  As argued by \citet{Dok03, Dok04}, our
follow-up spectroscopy is consistent with the $J_{\rm s} - K_{\rm s} \geq 2.3$
criterion selecting $z > 2$ galaxies with a high efficiency of $\sim 80\%$.

The $J_{\rm s} - K_{\rm s}$ colour cut at 2.3~mag is slightly bluer than used
in other recent studies \citep[in particular][]{Tot01, Sar03}.  Our objective
is not to focus on the rare extreme objects but to allow inclusion of systems
with rest-frame colours characteristic of normal nearby galaxies, which may
be missed by rest-frame UV-selected surveys.
Observed colours of $J_{\rm s} - K_{\rm s} \geq 2.3~{\rm mag}$ correspond
to rest-frame colours of $B - V \ga 0.5$ or $U - B \ga 0.1~{\rm mag}$ at
$z \approx 2.5$.  These rest-frame colours encompass those of the majority
of present-day luminous galaxies, from elliptical to intermediate-type
spiral galaxies \citep[e.g.][]{But94, Jan00}.

As discussed by \citet{Lab03}, contamination by foreground cool stars
is expected to be negligible since most known M, L, and T dwarfs have 
$J_{\rm s} - K_{\rm s}$ colours bluer than 2.3~mag; others such as
extreme carbon stars or Mira variables that could be redder are rare
and unlikely to be present in pencil beam surveys at high galactic
latitudes such as our FIRES fields \citep[see also][]{Hal01, Sar03}.
From NIR spectroscopy of six DRGs in the \mscl\ field and the Chandra
Deep Field South (CDF-S), \citet{Dok04} argued that strong emission
lines in the $K_{\rm s}$ band are unlikely to account for the observed
red colours of most of the population.  The main contaminants are more
likely very dusty ($A_{V} > 2~{\rm mag}$) galaxies at $1 < z < 2$,
probably at the $\sim 20\%$ level \citep{Dok03}.

Another potential concern is that samples selected with
the $J_{\rm s} - K_{\rm s} \geq 2.3$ criterion suffer from
contamination by AGN.  However, the majority of the bright DRGs we
identified in the HDF-S and \mscl\ fields (\S~\ref{Sub-sample}) are well
resolved in our $K_{\rm s}$-band imaging data, implying that their NIR
broad-band fluxes are not strongly influenced by continuum emission from an
AGN.  Emission lines from type~II AGN could also influence the NIR colours.
The rest-frame optical spectra of \citet{Dok04} show that this effect is
minor for the confirmed type~II AGN in our samples, and we are pursuing a
NIR spectroscopy program to confirm this result
\footnote{The fairly high fraction of $\sim 40\%$ of type~II AGN in the
spectroscopic sample \citep{Dok03, Dok04} is likely to result from the
spectroscopic selection bias towards galaxies that are bright in the
observer's optical and with bright emission lines.}.
For the moment, we conclude it is unlikely that emission from AGN
affects the NIR fluxes and colours in a substantial way.

\subsection{Sample Selection}
            \label{Sub-sample}

For the purpose of SED analysis and modeling, our goal was to construct a
sample of $J_{\rm s} - K_{\rm s}$ selected candidate $z \geq 2$ galaxies with 
robust NIR colours, reliable SEDs, and spanning a redshift range that would
allow consistent comparisons with the well-studied LBGs.  We required that the
$J_{\rm s} - K_{\rm s} \geq 2.3~{\rm mag}$ objects have a signal-to-noise ratio
$\rm S/N > 15$ on their $K_{\rm s}$-band colour flux and a minimum of 30\%
of the total exposure time (or relative weight $w \geq 0.3$) in all bands
\footnote{The weights in each band for the HDF-S are normalized to the
respective total exposure time over the entire field covered by the maps.
For the \mscl\ mosaic, the NIR ISAAC fields have somewhat different total
integration times and the highest overall weights are reached in the small
overlapping regions.
To ensure that the selection area covers fairly the mosaic, we considered the
weights for the NIR data normalized to the total exposure time in each of the
ISAAC fields individually.  For sources in the overlapping regions, we
required that $w \geq 0.3$ in at least one of the field.}.
This yielded initial samples of 14 objects at
$K_{\rm s}^{\rm tot} < 22.5~{\rm mag}$ in HDF-S and 31 at 
$K_{\rm s}^{\rm tot} < 21.7~{\rm mag}$ in the \mscl\ field.
The effective sky-plane areas are 4.48 and $\rm 23.86~arcmin^{2}$,
respectively.

\begin{figure}[p]
\epsscale{0.50}
\plotone{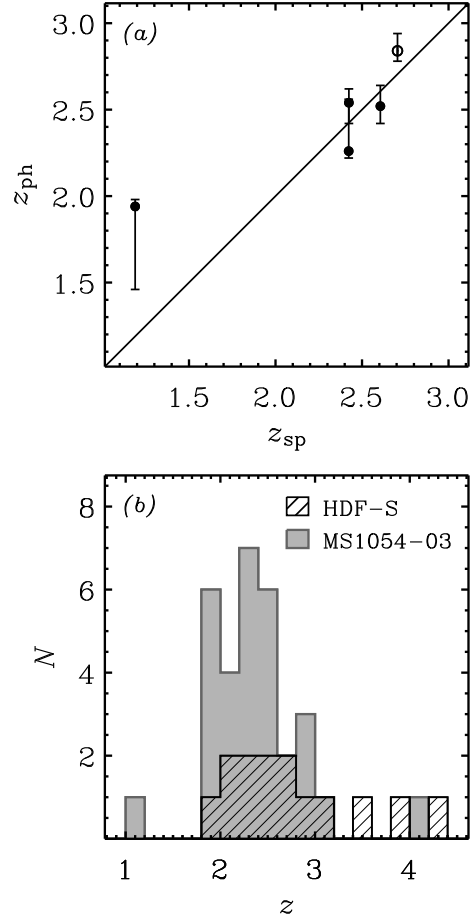}
\vspace{0.0cm}
\caption{
Redshift distributions of the $J_{\rm s} - K_{\rm s} \geq 2.3$
selected galaxies in the FIRES fields.  The sample includes
objects at $K_{\rm s}^{\rm tot} < 22.5~{\rm mag}$ from HDF-S
and $K_{\rm s}^{\rm tot} < 21.7~{\rm mag}$ from the
$\rm MS\,1054-03$ field.
({\em a\/}) Comparison between the photometric and spectroscopic
redshifts for the objects with $z_{\rm sp}$ determinations
(available in $\rm MS\,1054-03$ only).
The error bars represent the $1\sigma$ uncertainties on $z_{\rm ph}$.
The diagonal line indicates direct proportionality between
$z_{\rm ph}$ and $z_{\rm sp}$.
Object $\rm MS-140$ is also shown here ({\em open circle}) although it is not
formally part of the sample; it satisfies all the selection criteria except
the minimum exposure time in the NIR bands (see \S~\ref{Sub-sample}).
({\em b\/}) Redshift distributions of the HDF-S ({\em hatched histogram\/})
and $\rm MS\,1054-03$ ({\em grey-filled histogram\/}, excluding $\rm MS-140$)
samples.  When available, the spectroscopic redshift was used instead of
the photometric redshift.
\label{fig-z}
}
\end{figure}

The photometric redshifts of the selected sources lie in the range
$1.88 - 4.26$.  Spectroscopic redshifts are available for four of them in the
\mscl\ field (\citealt{Dok03, Dok04}; S. Wuyts \etal 2004, in preparation).
Three have $z_{\rm sp}$ between 2.42 and 2.61, and their
$z_{\rm ph}$ agrees with $z_{\rm sp}$ within $1\sigma$.
The fourth has $z_{\rm ph} = 1.94^{+0.04}_{-0.48}$ and turned out to be
an interloper at $z_{\rm sp} = 1.189$.  We note that the large uncertainty
towards low redshifts for this source reflects the presence of a long tail
of $z_{\rm ph}$ solutions with comparable likelihood.
Figure~\ref{fig-z} compares the $z_{\rm ph}$'s
with $z_{\rm sp}$'s and shows the redshift distributions of the samples,
where we used $z_{\rm sp}$ whenever available instead of $z_{\rm ph}$.
For the present study, we focused on the $J_{\rm s} - K_{\rm s}$ selected
sources with redshift between 2 and 3.5.  We chose this interval because
we are primarily interested in galaxies at $z \geq 2$ and it matches well
the range for LBGs selected from space-based HST WFPC2 data.  Ground-based
selected LBGs typically lie at somewhat higher redshift with less overlap
with the range covered by our DRG samples.  This is mainly because the
selection criteria rely on a different set of filters with, notably,
a redder ``$U$'' bandpass \citep[for a discussion, see, e.g.,][]{Gia01}.

The samples finally adopted, with $J_{\rm s} - K_{\rm s} \geq 2.3$
and $2 \leq z \leq 3.5$, consist of 11 objects
at $K_{\rm s}^{\rm tot} < 22.5~{\rm mag}$ in HDF-S and 23
at $K_{\rm s}^{\rm tot} < 21.7~{\rm mag}$ in the \mscl\ field.
The redshift distribution for the HDF-S (\mscl\ field) sample
has a mean $\langle z\rangle = 2.60$ (2.44),
median $\langle z\rangle_{\rm med} = 2.50$ (2.42),
and dispersion $\sigma(z) = 0.42$ (0.30).
These values are computed using the $z_{\rm sp}$ whenever available,
and are essentially unchanged when considering only the $z_{\rm ph}$'s
(the largest difference is $\langle z_{\rm ph}\rangle_{\rm med} = 2.30$
for the \mscl\ field).  There are four objects just below the $z = 2$
limit whose $1\sigma$ $z_{\rm ph}$ error would allow a $z > 2$ and,
conversely, five objects just above whose $1\sigma$ $z_{\rm ph}$ error
would allow $z < 2$; we verified that these small sample variations
would not alter the conclusions of our SED analysis and modeling.
Table~\ref{tab-sample} gives relevant characteristics of the adopted
DRG samples.

We note that not all the \mscl\ field objects for which spectroscopy
is presented by \citet{Dok03, Dok04} are included in our sample.
The initial selection for the spectroscopic follow-up was based on a
preliminary version of the reduced data and photometry.  In the final
photometric catalogue (v3.1b), the two galaxies with spectral signatures
indicating an AGN are slightly bluer than the $J_{\rm s} - K_{\rm s}$ cutoff
\citep[$\rm MS-1035$ and $\rm MS-1356$, numbered 1195 and 1458 in][]{Dok03}.
The bright $K_{\rm s}^{\rm tot} = 19.77~{\rm mag}$ and red
$J_{\rm s} - K_{\rm s} = 2.69 \pm 0.09~{\rm mag}$ object $\rm MS-140$
\citep[number 184 in][]{Dok03} satisfies all criteria except for the minimum
integration time in the NIR bands, with relative weights $w(J_{\rm s}) = 0.15$,
$w(H) = 0.21$, and $w(K_{\rm s}) = 0.18$.  However, it has a spectroscopic
redshift (included in Figure~\ref{fig-z}{\em a\/}) and, at $z_{\rm sp} = 2.705$,
lies in the range considered for our analysis.  $\rm MS-140$ constitutes a
valuable addition that we will discuss where relevant but keep out of the
formal sample.

\begin{figure}[p]
\epsscale{0.60}
\plotone{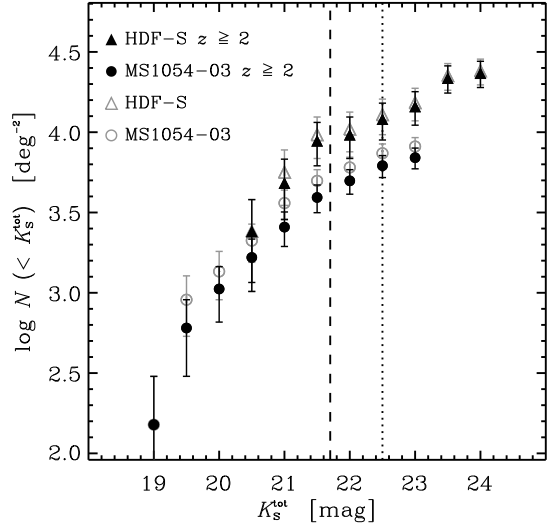}
\vspace{-1.0cm}
\caption{
Cumulative $K_{\rm s}$-band raw number counts of
$J_{\rm s} - K_{\rm s} \geq 2.3~{\rm mag}$ objects in the FIRES fields.
The counts for HDF-S ({\em triangles\/}) and the $\rm MS1054-03$ field
({\em circles\/}) include $J_{\rm s} - K_{\rm s}$ selected objects down
to the respective $5\sigma$ total limiting magnitudes for point sources
of $K_{\rm s}^{\rm tot} = 23.8$ and 23.1~mag.  The effective selection
areas are $\rm 4.48~arcmin^{2}$ for HDF-S and $\rm 23.86~arcmin^{2}$
for the \mscl\ field (see \S~\ref{Sub-sample}).
The counts for the $\rm MS1054-03$ field are not corrected for the lensing
effects of the cluster, which are small (on average $\approx 20\%$ in both
flux and area magnification) and to first order cancel out.
Filled symbols indicate counts for objects with redshift $z \geq 2$
while open symbols show counts with no redshift restriction
(the $z_{\rm sp}$ was used whenever available).
The error bars indicate the Poisson uncertainties.
The vertical lines show the magnitude cutoffs of the samples adopted for
analysis ensuring a $\rm S/N > 15$ on the $K_{\rm s}$-band colour flux,
at $K_{\rm s}^{\rm tot} = 22.5$ for HDF-S ({\em dotted line\/})
and 21.7~mag for $\rm MS\,1054-03$ ({\em dashed line\/}).
\label{fig-numcounts}
}
\end{figure}

Figure~\ref{fig-numcounts} shows the raw cumulative number counts per unit
area as a function of total $K_{\rm s}$ magnitude of the DRGs, for both cases
of $z \geq 2$ and of no redshift restriction.  To assess the variations of
surface density with limiting magnitude and between the two fields, the
counts are extended to the $5\sigma$ $K_{\rm s}^{\rm tot}$ limits for
point sources of 23.8~mag in HDF-S and 23.1~mag in the \mscl\ field.
We did not account here for the lensing by the \mscl\ cluster.  The
corrections would be small, with an average dimming by $\rm \approx 0.2~mag$
and a mean reduction by $\approx 20\%$ from image- to source-plane area
for the redshift range covered by the sources.
These effects are opposite so that to first order, the surface
density to a given apparent magnitude is insensitive to lensing (the exact
effect depends on the slope of the luminosity function, which is unknown).

Per unit area, the \mscl\
field is richer in sources brighter than $K_{\rm s}^{\rm tot} = 20~{\rm mag}$
while HDF-S has higher surface densities for fainter limits down to 23~mag
(where the completeness levels still are $\ga 90\%$ for both fields).  These
differences are not due to the lensing magnification by \mscl.  For $z \geq 2$,
there are five DRGs at $K_{\rm s}^{\rm tot} < 20~{\rm mag}$ in the \mscl\ field
and none in HDF-S; at fainter limits, HDF-S has higher surface densities by a
factor of two on average.  The redshift cut has little impact: without it,
there are seven DRGs at $K_{\rm s}^{\rm tot} < 20~{\rm mag}$ in the \mscl\
field and still none in HDF-S, and the surface densities at fainter limits are
1.7 times higher on average in HDF-S.  The two fields cover small areas and
cosmic variance could easily account for a large part of the differences in
the surface densities, the more so in view of the possible strong clustering
of the DRG population \citep{Dad03}.  This might also explain the deficiency
in HDF-North, with similar survey area as HDF-S, where we find no source with
$J_{\rm s} - K_{\rm s} \geq 2.3$ colours at magnitudes brighter than
$K_{\rm s} = 21~{\rm mag}$ using ground-based photometry \citep[from][]{Fer99}
\footnote{We consider here specifically colours involving the ground-based
$J$ band because the corresponding F110W bandpass of NICMOS onboard HST is
significantly bluer and about twice as broad as the $J$ or $J_{\rm s}$
bandpasses (the colour term is negligible between the latter two).
Consequently, F110W photometry is not useful to select galaxies red in their
$J_{\rm s} - K_{\rm s}$ colours.}.

\begin{figure}[p]
\epsscale{0.60}
\plotone{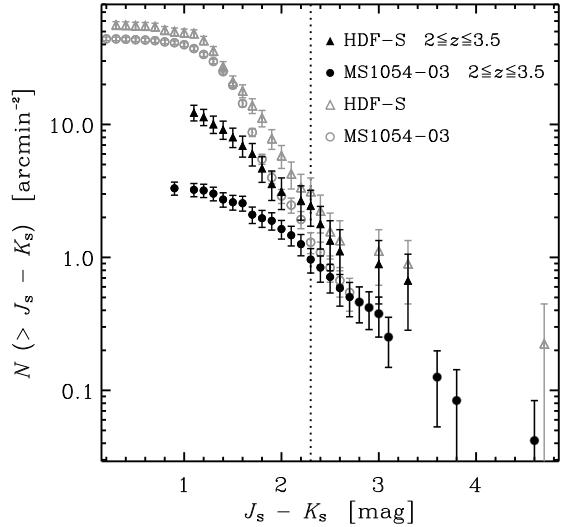}
\vspace{-1.0cm}
\caption{
Cumulative raw number counts as a function of $J_{\rm s} - K_{\rm s}$
colour cutoff for $K_{\rm s}$-band selected sources in the FIRES fields.
The counts are given as number of objects per unit area $N$ with colour
redder than the $J_{\rm s} - K_{\rm s}$ of the abscissa (the counts for
the $\rm MS\,1054-03$ field are uncorrected for the area magnification
by the cluster, a small effect of $\approx 20\%$ on average).
The counts for HDF-S ({\em triangles\/}) and the $\rm MS1054-03$ field
({\em circles\/}) include objects down to $K_{\rm s}^{\rm tot} = 22.5$
and 21.7~mag, and over the effective selection areas of 4.48 and
$\rm 23.86~arcmin^{2}$, respectively (see \S~\ref{Sub-sample}).
Filled symbols indicate counts for objects with redshift $2 \leq z \leq 3.5$
while open symbols show counts with no redshift restriction (the $z_{\rm sp}$
was used whenever available). 
The error bars indicate the Poisson uncertainties.
The vertical dotted line shows the $J_{\rm s} - K_{\rm s}$ colour cutoff at
2.3~mag.  Most galaxies with $J_{\rm s} - K_{\rm s} \geq 2.3$ colours have
redshifts at $z > 2$.
\label{fig-numcounts_col}
}
\end{figure}

Figure~\ref{fig-numcounts_col} plots the raw cumulative counts as a
function of $J_{\rm s} - K_{\rm s}$ cutoff for all $K_{\rm s}$-band
selected sources at $K_{\rm s}^{\rm tot} < 22.5~{\rm mag}$ in HDF-S and
$\rm < 21.7~mag$ in the \mscl\ field (the magnitude limits ensuring
$\rm S/N > 15$ on the $K_{\rm s}$-band colour fluxes of the DRGs).
For $2 \leq z \leq 3.5$, the DRG samples are not strongly sensitive to the
exact colour cut: varying the criterion between 2.2 and 2.4~mag implies 
$11^{+1}_{-3}$ objects for HDF-S and $23^{+7}_{-3}$ for the \mscl\ field,
or $10\%-30\%$ variations.  Imposing no redshift limits, the same colour
cut variations result in $14^{+1}_{-4}$ objects for HDF-S and $31^{+15}_{-5}$
for the \mscl\ field; the fraction of galaxies at $z < 2$ increases rapidly
for $J_{\rm s} - K_{\rm s}$ colours bluer than 2.3~mag.
The median $1\sigma$ uncertainty on the $J_{\rm s} - K_{\rm s}$ colours among
the adopted DRG samples is 0.13~mag and all those with uncertainties larger
than 0.2~mag (the maximum is 0.43~mag) have colours redder than 3.0~mag.

\subsection{Comparison LBG Sample}
            \label{Sub-sample1}

We also constructed a sample of LBGs taken in HDF-S.  We used
the original HDF-S HST WFPC2 photometric catalogue from \citet{Cas00}
and selected the LBGs using the criteria introduced by \citet{Gia01}.
Of the 139 selected LBGs at $V_{\rm 606,\,AB} < 27~{\rm mag}$, 91 are
cross-identified in our FIRES $K_{\rm s}$-band selected catalogue
and have $w \geq 0.3$ in our full set of maps.  All LBGs in this
sub-sample are brighter than $K_{\rm s}^{\rm tot} \leq 24.0~{\rm mag}$.
To make up the comparison LBG sample, we then selected those satisfying
the same $K_{\rm s}^{\rm tot}$ limit as for the HDF-S DRGs and with
redshift $2 \leq z \leq 3.5$.  The requirement on the $K_{\rm s}$-band
magnitudes ensures that the selected LBGs have reliable NIR photometry
and rest-frame $V$-band luminosities similar to those of the selected DRGs.
The adopted sample consists of 33 LBGs
at $K_{\rm s}^{\rm tot} < 22.5~{\rm mag}$.  Their redshift distribution
has $\langle z \rangle = 2.68$, $\langle z \rangle_{\rm med} = 2.76$,
and $\sigma(z) = 0.42$.  The mean and median redshifts are close to those
of the $2 \leq z \leq 3.5$ DRGs, especially from HDF-S.  The average
(and median) $J_{\rm s} - K_{\rm s}$ colour is 1.68~mag with dispersion
$\sigma = 0.28~{\rm mag}$, and only two of these LBGs are redder than
$J_{\rm s} - K_{\rm s} = 2.3~{\rm mag}$.

For this LBG sample, we will hereafter use the corresponding
$U_{300}\,B_{450}\,V_{606}\,I_{814}\,J_{\rm s}\,H\,K_{\rm s}$ photometry from
the FIRES HDF-S catalogue (v3.0e).  Although selected at different wavelengths,
the use of SEDs from the same photometric catalogue ensures uniformity in the
data sets for both DRGs and LBGs and more directly comparable results in the
analysis and modeling.  The requirement that our LBG sample be also
detected in the $K_{\rm s}$ band makes our selection slightly different
than the classical purely optical selection.
We did not construct a similar LBG sample for the \mscl\ field because of
its ground-based $UBV$ data, which would select LBGs in a somewhat higher
redshift range than spanned by the DRG samples.

\section{PHOTOMETRIC PROPERTIES OF THE SAMPLES}
            \label{Sect-prop}

The large numbers of DRGs that we identified from 
$J_{\rm s} - K_{\rm s} \geq 2.3~{\rm mag}$ in the FIRES survey suggest they
may represent a significant component of the high-redshift galaxy population
\citep{Fra03, Dok03}.  It is thus important to understand their nature,
in particular their stellar populations and dust content.
Before undertaking the detailed theoretical modeling, we here examine the
broad-band properties of the sample to assess the relative importance of
aging and extinction in a qualitative but less model-dependent manner.

\subsection{Observed Optical to Near-infrared SEDs}
            \label{Sub-SEDs_obs}

The observed SEDs of DRGs in both FIRES fields exhibit a wide variety in
their shapes, especially in the optical regime.  Figure~\ref{fig-SEDs}
presents the SEDs of selected representative objects in each of the HDF-S
and \mscl\ field.  The galaxies are sorted according to increasing
$I_{814} - K_{\rm s}$ colour, and the SEDs are normalized to the flux
density in the $K_{\rm s}$ band (the overplotted SED fits will be 
discussed in \S~\ref{Sect-models}).
At one extreme, some of the DRGs appear overall very red with little if any
flux in the optical bands.  At the other, some have blue observed optical
colours indicative of on-going star formation activity.  Most exhibit a rather
abrupt transition redwards of the $J_{\rm s}$ band reminiscent of a spectral
break in the continuum.  A few show a rather smooth and gradual increase
in flux with wavelength suggestive of a featureless continuum reddened
by dust obscuration.

\subsection{Rest-frame SEDs Properties}
            \label{Sub-SEDs_rf}

The rest-frame optical colours of the DRGs lie within the range covered by
present-day normal galaxies, in striking contrast with the much bluer LBGs.
Figure~\ref{fig-rfSEDs} shows the rest-frame SEDs of our samples of DRGs and
LBGs at $2 \leq z \leq 3.5$.  All SEDs are dereddened for Galactic extinction.
We accounted for the average intergalactic attenuation due to line blanketing
by Ly$\alpha$ ($\langle D_{A} \rangle$) and higher-order Lyman lines
($\langle D_{B} \rangle$) following the prescriptions of \citet{Mad95}.
The SEDs are compared with the empirical UV-optical template spectra
of nearby galaxies of types E, Sbc, Scd, and Im from \citet*{Col80}.
The spectra shown here were extended beyond their original
$\lambda = 1400 - 10000$~\AA\ coverage by \citet*{Bol00} using the 
evolutionary synthesis code of \citet[][GISSEL98 version]{Bru93}
with appropriate model assumptions.
The SEDs and spectral templates are normalized to a flux density 
$f_{\lambda}$ of unity at $\lambda_{\rm rest} = 2500$~\AA\ using
linear interpolation between the adjacent rest-frame data points.

At $\lambda_{\rm rest} \ga 2500$~\AA, the SEDs of all DRGs fall within
the envelope of normal galaxies.  None of the DRGs has rest-frame optical
colours as red as the earliest-type nearby ellipticals, perhaps not
surprisingly so because the universe is only a few times $\rm 10^{9}~yr$
old at $2 \leq z \leq 3.5$.  At shorter wavelengths, the SEDs span a wide
range in rest-frame UV slopes encompassing those of normal galaxies and
extending to much redder slopes, suggesting a range in level of star
formation activity and in interstellar extinction.
In contrast, the LBG SEDs exhibit less scatter especially in the UV and
are systematically bluer than late-type spirals, as first demonstrated by
\citet{Pap01}.  These authors showed that the SEDs of their HDF-N LBGs
fit well in the envelope defined by the local starburst templates of
\citet{Kin96} and are consistent with the presence of a young and
modestly obscured stellar population.

Figure~\ref{fig-rfSEDs_sim} illustrates the effects of extinction, where we
used the Im-type template of \citet{Col80} and applied the extinction law of
\citet{Cal00} assuming a simple uniform foreground screen of obscuring dust.
The template SEDs inherently include some intrinsic reddening so that our
simulations represent additional amounts of extinction.  Interestingly enough,
the Im template combined with added extinction of $A_{V}$ between 0 and 3~mag
brackets the range of rest-frame SED shapes of our DRG sample.
Large extinction offers a natural interpretation for the many objects with
very red rest-frame UV emission and consequently could also account in part
for the red rest-frame optical colours.  
Figures~\ref{fig-rfSEDs} and \ref{fig-rfSEDs_sim} do not allow us to
assess the relative importance of evolved stellar populations and
extinction effects in producing the observed SEDs of DRGs.  However,
they demonstrate clearly that DRGs and LBGs are distinct in their
rest-frame UV-optical SEDs, presumably because of very different
dominant stellar populations and/or extinction levels.

In Figure~\ref{fig-rfSEDs}, the SEDs of the three spectroscopically-confirmed
DRGs among our sample along with $\rm MS-140$ (see \S~\ref{Sub-sample})
are plotted with different symbols.  The comparison indicates that the galaxies
in the spectroscopic sample do not deviate strongly in their rest-frame SEDs
from the full photometric sample.  On the other hand, it also shows that the
spectroscopic sample does not probe the entire range of properties seen
for the ensemble of DRGs.

\subsection{Optical Break versus Dust Extinction}
            \label{Sub-NIRcol}

The wavelength sampling and resolution of our data are too coarse to
delineate accurately a Balmer/4000\,\AA\ break.  Even when strongest, the
break represents a moderate flux jump by a factor of $\approx 2$, a subtle
feature compared to the Lyman break at 912~\AA.  For the brightest DRG of
the spectroscopic sample, the presence of a break is confirmed by the
detection of a significant drop in continuum level in its NIR spectrum
collected with NIRSPEC at the Keck Telescope \citep{Dok04}.
Here, we examine the observed NIR properties of our ensemble of DRGs using
diagnostic colour-colour diagrams, which allow us to establish that a break
is required to explain the colours of an important fraction of them.

Specifically, we test whether the red observed NIR colours of DRGs
(1) can be reconciled with a very young ($\rm \la 10^{8}~yr$)
stellar population that is highly obscured by dust, or if
(2) the presence of a more evolved ($\rm \ga 10^{8}~yr$)
stellar population whose SED has a prominent Balmer/4000\,\AA\
break is necessary.
In exploring various colour combinations, we relied on the sharpness of the
Balmer/4000\,\AA\ break versus the smoother wavelength dependence of dust
attenuation.  Extinction makes all colours redder while a prominent break
produces a larger contrast between colours because of the steep drop in flux
at the break.  As a stellar population ages, the continuum slope intrinsically
reddens because cooler stars dominate the optical light but the break remains
comparatively sharp.  For redshifted populations, these differences in colour
contrast will be reflected in the variations with $z$ in colour-colour space.

We found that $H - K_{\rm s}$ versus $J_{\rm s} - H$ and $J_{\rm s} - H$
versus $I_{814} - J_{\rm s}$ provide the most sensitive diagnostic diagrams.
The use of these adjacent bandpasses traces as closely as possible the
Balmer/4000\,\AA\ break in the range $z \approx 1 - 3.5$ while minimizing
reddening due to dust and to cool dominant stars.
Figure~\ref{fig-NIRcol} compares in these diagrams the colour distributions
of the DRGs with the redshift evolution at $2 \leq z \leq 4$ of selected model
SEDs: 250~Myr and 1~Gyr old single-age stellar populations (SSPs), a passively
evolving population formed at $z_{\rm f} = 5$, and populations with constant
star formation rate (CSF) and ages of 10~Myr, 100~Myr, and 1~Gyr.
We computed the synthetic spectra and colours using the codes of
\citet{Bru03} and \citet{Bol00} as for the SED modeling and with the same
standard assumptions (see \S~\ref{Sect-models}).  We also plot tracks for the
100~Myr old CSF model with $A_{V}$ in the range $\rm 1 - 6~mag$, applying the
extinction law of \citet{Cal00} and a uniform foreground screen geometry.

The model tracks in the top panels of Figure~\ref{fig-NIRcol} show the expected
trend that the older the age, the larger the amplitude of the colour variations
with $z$ as the stronger break moves redwards across the successive bandpasses.
The effect is maximized for SSPs or passively evolving populations, where the
break is not diluted by the light from younger stars with shallower or no
absorption features and with bluer continuum.  CSF models are consistently
bluer in all colours and occupy narrower regions than SSPs of the same age.
Extinction effects shift the tracks along paths that run diagonally across the
diagrams but preserve their amplitude at fixed age.  As shown schematically
in the bottom panels, the locus of CSF models with varying extinction gets 
wider with older ages, because of the stronger break.  The effects of aging
and extinction can be distinguished in that the loci of dusty model tracks
broaden with increasing age in the direction perpendicular to the extinction
path, the key feature we here exploit.

The immediate result from these diagrams is that the colours of the ensemble
of DRGs cannot be reconciled with very young and highly obscured populations.
Dusty CSF models need the broadening due to significant aging to
better match the observed spread perpendicular to the extinction vector.
To quantify the fraction of DRGs not overlapping with the dusty models,
Figure~\ref{fig-NIRcol_proj} presents histograms where the colour
distributions have been rebinned as a function of the deviation in
$J_{\rm s} - H$ and $I_{814} - J_{\rm s}$ around the line best representing
the dusty 100~Myr old CSF tracks (using robust linear regression).  The maximum
extent of the tracks about this line is indicated in the plots.  The central
loci and maximum widths for the 10~Myr and 1~Gyr old dusty CSF models are
plotted as well.  About two-thirds of the DRG sample has colours which are
inconsistent with those of a 10~Myr old obscured star-forming population.
The fraction drops to $30\% - 55\%$ for an age of 100~Myr, and $5\% - 25\%$
for 1~Gyr (depending on the colour considered).

Interestingly, the $H - K_{\rm s}$ versus $J_{\rm s} - H$ diagram shows that
if DRGs are actively star-forming, then the $J_{\rm s} - K_{\rm s} \geq 2.3$
colour cut will miss even those with a strong optical break unless they are
highly obscured.  In other words, the criterion preferentially selects not
only for the most evolved star-forming systems but for the most obscured
ones as well.  The need for high extinction is alleviated for scenarios
where the star formation rate is declining with time (as we discuss further in
\S~\ref{Sect-models}), but the $J_{\rm s} - K_{\rm s}$ colour cut still selects
for evolved galaxies.  Figures~\ref{fig-NIRcol} and \ref{fig-NIRcol_proj}
demonstrate that the red NIR colours of the ensemble of DRGs cannot be
attributed to extinction effects alone.  For a significant fraction of the
objects, the colours require the presence of a strong Balmer/4000\,\AA\ break
produced by stars with ages $\rm \ga 250~Myr$.

\section{MODELING OF THE SPECTRAL ENERGY DISTRIBUTIONS}
	 \label{Sect-models}

In this section, we describe our model ingredients and fitting methodology.
We present the results for our set of ``standard'' parameters with fixed
initial mass function (IMF), metallicity, extinction law, and two choices
of star formation history (SFH).  We also test the sensitivity of the
results to changes in the SFH, metallicity, and extinction law.
We modeled the broad-band SEDs of our DRG sample following the
approach adopted in recent similar studies of high redshift
galaxies \citep[e.g.][]{Pap01, Sha01, Dok04}.  
We fitted redshifted synthetic spectra of stellar populations to the
SEDs to constrain simultaneously the age and extinction, keeping $z$
fixed for each object.  We derived the stellar masses ($M_{\star}$)
and star formation rates (SFRs) by scaling those of the best-fit input
model so as to reproduce the observed fluxes.  We determined the
stellar mass-to-light (M/L) ratios using the attenuated rest-frame
$V$-band luminosities ($L_{V,\star}$) computed as described by 
\citet[][their Appendix C]{Rud03}.  Although the extrinsic
(i.e. dust-attenuated) luminosities from our best-fit models agree
to better than 10\% on average, we preferred the less model-dependent
estimates of \citet{Rud03}.

\subsection{Model Ingredients}    \label{Sub-model_ingr}

We generated the synthetic spectral templates with the most recent version
of the evolutionary synthesis code developed by G. Bruzual and S. Charlot
\citep{Bru03}.  This ``BC03'' version features improved stellar
evolutionary tracks with observationally motivated prescriptions for the
thermally-pulsing asymptotic giant branch phase.  It also incorporates
updated and newly available stellar libraries for two sets of spectral
resolution and coverage.  We used the lower resolution set covering
$\lambda = 9$~\AA\ to 160~\micron\ relying on the BaSeL~3.1 library.
We selected the ``Padova 1994'' evolutionary tracks, which are preferred by
\citeauthor{Bru03} over the more recent ``Padova 2000'' tracks because the
latter may be less reliable and predict a hotter red giant branch leading to
worse agreement with observed galaxy colours.  We used the solar metallicity
set of tracks.  The only constraints available for the chemical abundances
of DRGs come from the [\ion{N}{2}]$\lambda 6584$\,\AA/H$\alpha$ line ratio
measured in two of the galaxies observed spectroscopically, which suggests
high metallicities of $Z \sim 1 - 1.5~{\rm Z_{\odot}}$ \citep{Dok04}.
These DRGs appear more metal-rich than the five LBGs at $z \sim 3$ studied
by \citet[][$0.1 - 0.5~Z_{\odot}$]{Pet01} and similar to the seven UV-selected
star-forming ``BX/MD'' objects at $z \sim 2$ for which \citet{Sha04} inferred
solar, and possibly super-solar, metallicities.  In all cases, however, the
determinations rely on limited samples and suffer from large uncertainties.

The computed grid of spectral templates covers ages between $10^{5}$ and
$\rm 2 \times 10^{10}~yr$.  The SFH most appropriate for our objects is
unknown and broad-band photometry alone is not effective at constraining
it \citep[see \S~\ref{Sub-paramvar}; also, e.g.,][]{Pap01, Sha01}.
We considered two star formation rate parametrizations: constant in
time with $R(t) = R(t = 0) \equiv R_{0}$ and exponentially declining in
time as $R(t) = R_{0} \exp(-t/\tau)$ with timescale $\rm \tau = 300~Myr$
(hereafter ``CSF'' and ``\tauint'' models).   These exact choices are
somewhat arbitrary.  The CSF model is intended to represent the case of
significant on-going or recent star formation activity.
Evidence for this scenario in part of the DRG population includes the
detection of H$\alpha$ and/or [\ion{O}{3}]\,$\lambda\lambda 4959,5007$~\AA\
lines in the non-AGN sources of the spectroscopic sample \citep{Dok04} and
the blue rest-frame UV SED of several others among the full sample (see
Figures~\ref{fig-SEDs} and \ref{fig-rfSEDs}).  The \tauint\ model allows
for the possibility of quiescent systems that underwent a period of
enhanced star formation in their past.

We adopted a Salpeter IMF (${\rm d}N/{\rm d}m \propto m^{-2.35}$) between 0.1
and 100\masssol.  The IMF is unconstrained for our objects, and a steep rise
down to the lower mass cutoff likely is unrealistic in view of the turnover
below 1\masssol\ inferred for the local IMF \citep[e.g.,][]{Kro01, Cha03}.
The most critical impact is on the derived masses and mass-to-light ratios
($M/L$), which depend strongly on the shape and cutoff of the low-mass IMF.
For instance, for a fixed rest-frame $V$-band luminosity, the derived masses
and M/L ratios would be about a factor of two lower with the \citet{Kro01}
or \citet{Cha03} IMFs (e.g., Figure 4 of \citealt{Bru03}; see also
\citealt{Pap01} for a discussion of the effects of different IMFs
in their SED modeling of LBGs).

We have used the mass of stars still alive $M_{\star}$
instead of the total mass of stars formed $M_{\rm tot}$.  $M_{\rm tot}$
is simply the integral of the SFR and corresponds to the total gas mass
consumed since the onset of star formation.  $M_{\star}$ is computed by
subtracting from $M_{\rm tot}$ the mass returned to the ISM by evolved stars
via stellar winds and supernova explosions \citep[see][for details]{Bru03}.
We preferred $M_{\star}$ in our analysis because it represents the mass
of the stars producing the light.  For the typical ages $\rm 1 - 2~Gyr$ of
interest here and with the adopted IMF, the cumulative mass-loss effects are
modest and comparable for the CSF and \tauint\ models, reaching $20\% - 25\%$
of $M_{\rm tot}$.  Even in the limit of an SSP where the effects are largest,
they amount to a similar fraction of $23\% - 26\%$.

Both interstellar extinction by dust within the objects and attenuation
due to intergalactic H opacity were applied to the BC03 templates while
performing the fits.
We explored a range of extinction with $0 < A_{V} < 3~{\rm mag}$.
Higher $A_{V}$ values are physically possible but unlikely to be actually 
measured from rest-frame UV/optical data because the corresponding
optical depths $\tau_{\lambda} = A_{\lambda} / 1.086$ largely exceed unity.
The extinction law and the geometry for the obscuring dust and emitting
sources are other uncertain parameters.  We adopted the law of \citet{Cal00}
derived empirically from observations of local UV-bright starburst galaxies
under the formalism of a foreground screen of obscuring dust.  As emphasized
by these authors, this extinction law and model geometry may not be valid in
dust-rich systems where the optical depths are large and the sources and
dust are more or less mixed spatially.
The average Lyman line absorption by the intergalactic medium (IGM) was
accounted for based on the prescriptions of \citet{Mad95}.  Lyman continuum
absorption was approximated by setting the flux of the templates equal to
zero at $\lambda_{\rm rest} < 912$~\AA.

\subsection{Fitting Procedure}    \label{Sub-model_proc}

We performed the fits using the publicly available HYPERZ photometric
redshift code, version 1.1 \citep{Bol00}.  We fixed the redshift to the
$z_{\rm ph}$ determined independently (see \S~\ref{Sect-data}) or to
the $z_{\rm sp}$ when available.  The code was allowed to fit simultaneously
the spectral template (age) and the extinction, for each of the SFH adopted.
Interstellar extinction is first applied to the synthetic spectra, which are
then redshifted and further attenuated for IGM absorption.
The fitting algorithm
relies on least-squares minimization, with the chi-squared calculated as
$\chi^2 = \sum_{n} [\,(f_{\rm obs} - b\,f_{\rm temp}(z))\,/\,\sigma\,]^2$.
The sum is performed over the $n$ photometric bands.  The template fluxes
$f_{\rm temp}(z)$ are computed from the reddened, redshifted, IGM-absorbed
template spectra using the total effective transmission functions in each
bandpass.  The scaling factor $b$ is determined from the full SED constructed
from the colour fluxes, corrected by the same amount to the total aperture
based on the $K$-band measurements; all SEDs have also been corrected for
Galactic extinction and, for the \mscl\ field sample, for the lensing
magnification prior to the fitting (\S~\ref{Sect-data}).

We treated any flux with $f_{\rm obs} < 2\sigma$ as a non-detection,
with upper limit equal to the $1\sigma$ uncertainty.
The assigned $1\sigma$ uncertainties and upper limits correspond to those
of the flux measurements.  We assumed a minimum error of 0.05~mag on the
photometry to avoid the fits being driven by a few data points with very
small errors.  This also accounts for uncertainties in the absolute flux
calibration.  Our results are not strongly influenced by this assumption.
If we do not impose a minimum photometric error, the median of the best-fit
extinction values remains the same for the two fields and two standard SFHs
considered.  The median $M_{\star}$'s vary by $\leq 1\%$, and the median SFRs
by $< 15\%$.  The median of the best-fit ages decreases from 1.7 to 1.0~Gyr
for the HDF-S DRGs and CSF, and increases from 1.0 to 1.4~Gyr for the \mscl\
field DRGs and the \tauint\ model, not affecting our findings that DRGs
harbour old stellar populations.

HYPERZ resamples the age grid from the BC03 models from 221 to 51 fixed
steps.  We required that the template ages considered for the fitting
do not exceed that of the universe at the redshift of each source.
We emphasize that the best-fit age, which we will denote $t_{\rm sf}$, is
strictly speaking the time elapsed since the onset of star formation.
An alternative, perhaps more meaningful definition is the SFR-weighted mean
age $\langle t \rangle_{\rm SFR}$ corresponding more closely to the age of
the stars contributing the bulk of the stellar mass.
For CSF models, $\langle t \rangle_{\rm SFR} = 0.5\,t_{\rm sf}$.
For exponentially decaying models,
$\langle t \rangle_{\rm SFR} = 
[t_{\rm sf} - \tau + \tau\,\exp(-t_{\rm sf}/\tau)]  /  
[1 - \exp(-t_{\rm sf}/\tau)]$,
which becomes $\langle t \rangle_{\rm SFR} \approx t_{\rm sf} - \tau$
when $t_{\rm sf} \gg \tau$.
Moreover, $t_{\rm sf}$ should not be over-interpreted as the age of the
entire galaxy.  It is important to remember that all the properties derived
from the SED modeling apply only to the stellar population that dominates
the fluxes in the observed wavelength range.  This does not preclude, e.g.,
the presence of a pre-existing stellar population that has faded out too
much \citep[a scenario explored for LBGs by][]{Pap01} or of a younger
population that is too obscured to be detected by our observations.

We estimated the confidence intervals for individual objects from
Monte-Carlo simulations.  For each source, we performed 500 synthetic 
realizations of the data by varying the fluxes by an amount randomly drawn
from the (Gaussian) distribution of the measurement uncertainties (including
the minimum error of 0.05~mag where appropriate).  For sources with only a
$z_{\rm ph}$ determination, we also varied the redshift by drawing randomly
from the associated probability distribution $P(z)$.  We then repeated the
fitting procedure for each realization.  We determined the 68\% confidence
intervals from the distributions of best-fit values obtained with the
simulated data.  The behaviour of the confidence intervals for the DRG
samples studied here are very similar to those of the spectroscopic sample
of \citet{Dok04} and of the LBGs modeled by \citet{Pap01} and \citet{Sha01},
i.e. often asymmetric around the best-fit values, reflecting strong
degeneracies notably between $t_{\rm sf}$ and $A_{V}$, and implying that
$M_{\star}$ is the best constrained quantity for the majority of objects;
we refer to the above papers for a more detailed discussion
\footnote{For the combined DRG sample, the median of the uncertainties given
by the 68\% confidence intervals for the CSF (\tauint) standard model are
$^{+17\%}_{-47\%}$ ($^{+18\%}_{-17\%}$) for the age $t_{\rm sf}$,
$^{+26\%}_{-24\%}$ ($^{+28\%}_{-22\%}$) for $M_{\star}$,
$^{+14\%}_{-28\%}$ ($^{+21\%}_{-27\%}$) for $M_{\star}/L_{V, \star}$, and
$^{+60\%}_{-19\%}$ (50\% to a factor of 2) for the instantaneous SFR.
For the HDF-S LBG sample also modeled in this work (\S~\ref{Sub-DRGs_LBGs}),
the uncertainties are
$^{+41\%}_{-29\%}$ ($^{+17\%}_{-29\%}$) for the age $t_{\rm sf}$,
$^{+27\%}_{-25\%}$ ($^{+13\%}_{-20\%}$) for $M_{\star}$,
$^{+22\%}_{-18\%}$ ($\pm 5\%$) for $M_{\star}/L_{V, \star}$, and
$^{+13\%}_{-22\%}$ ($^{+48\%}_{-41\%}$) for the instantaneous SFR.
For DRGs and LBGs, the uncertainties on $A_{V}$ are $\leq 0.3~{\rm mag}$
with both the CSF and \tauint\ models.}.
The derived uncertainties are smaller or comparable to the differences in
best-fit values obtained by varying the assumed SFH, metallicity, or extinction
law (see \S~\ref{Sub-paramvar}).  The main goal of this work is to characterize
the ensemble properties of DRGs (from the median and the global distributions
of best-fit values), and accounting for the uncertainties of individual sources
does not alter our general conclusions.  We will therefore not discuss them
further, except where most relevant.

\subsection{Modeling Results for the Standard Model Sets}
           \label{Sub-model_res}

\begin{figure}[p]
\epsscale{1.0}
\plotone{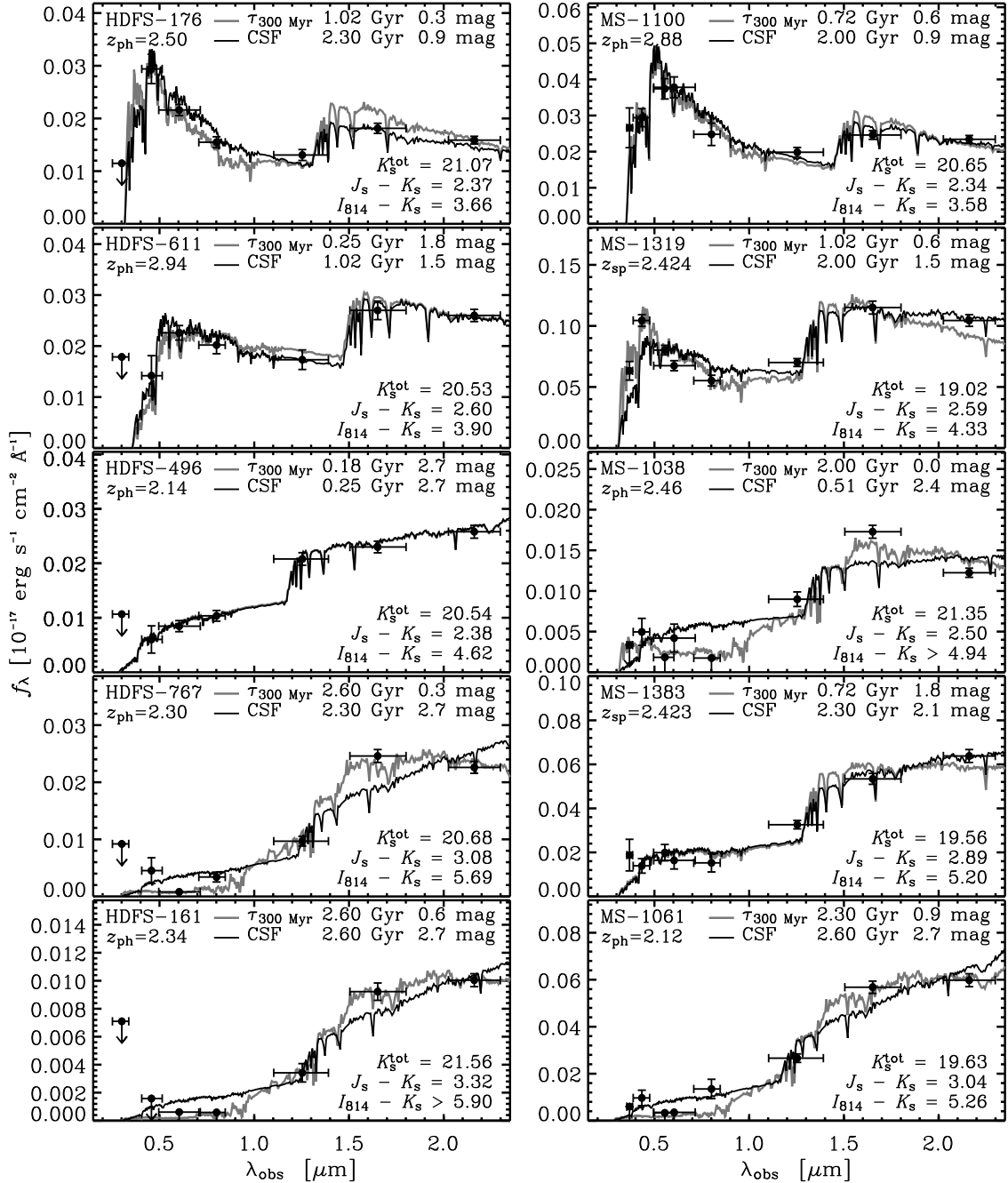}
\vspace{0.5cm}
\caption{
Broad-band SEDs of selected $J_{\rm s} - K_{\rm s} \geq 2.3$
objects at $2 \leq z \leq 3.5$ in the FIRES fields. 
The panels to the left show SEDs of HDF-S objects and to the right, SEDs
of objects in the $\rm MS\,1054-03$ field.  The galaxies are sorted by
increasing $I_{814} - K_{\rm s}$ colour.  All SEDs are corrected for
Galactic extinction.  The vertical error bars and
the upper limits represent the $1\sigma$ uncertainties on the flux
measurements or the minimum error of 0.05~mag assumed for the modeling
(see \S~\ref{Sub-model_proc}).  The horizontal error bars indicate
the bandpass widths at half the maximum transmission.
The best model fits for the two ``standard'' star formation
histories (see \S~\ref{Sub-model_ingr}) are overplotted:
a constant star formation rate (CSF; {\em black lines\/})
and an exponentially declining star formation rate with
$\rm \tau = 300~Myr$ ($\tau_{\rm 300\,Myr}$; {\em grey lines\/}).
The best-fitting age and visual extinction $A_V$ are given
for both models in each panel.
The models were computed for a Salpeter IMF between 0.1 and
$100~{\rm M_{\odot}}$, solar metallicity, and the \citet{Cal00}
extinction law.
\label{fig-SEDs}
}
\end{figure}

\begin{figure}[p]
\epsscale{0.80}
\plotone{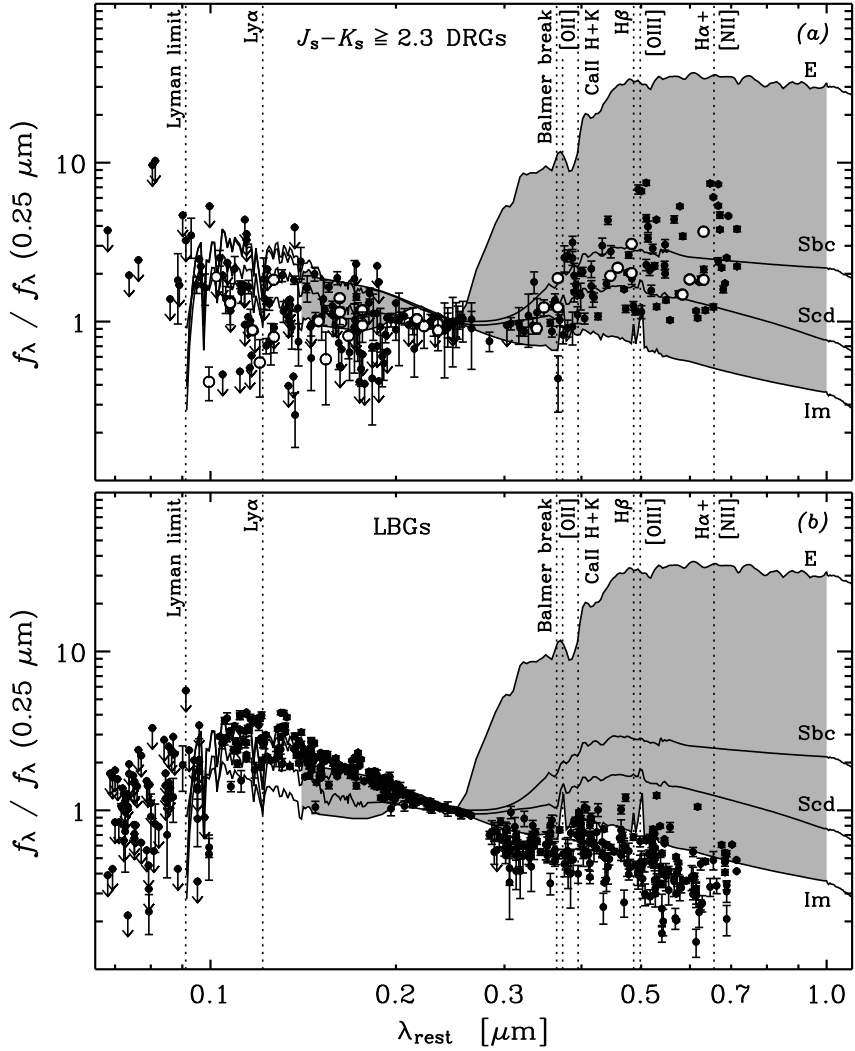}
\vspace{0.0cm}
\caption{
Rest-frame broad-band SEDs of $2 \leq z \leq 3.5$
galaxies in the FIRES fields.
({\em a\/}) $J_{\rm s} - K_{\rm s} \geq 2.3$ selected
objects at $K_{\rm s}^{\rm tot} < 22.5~{\rm mag}$ in HDF-S and
$K_{\rm s}^{\rm tot} < 21.7~{\rm mag}$ in the $\rm MS\,1054-03$ field.
The objects with available spectroscopic redshift are plotted with large
white-filled dots (including $\rm MS-140$, see \S~\ref{Sub-sample}).
({\em b\/}) Optically-selected LBGs at
$K_{\rm s}^{\rm tot} < 22.5~{\rm mag}$ in HDF-S.
All SEDs are corrected for Galactic extinction, and for the average
absorption due to intergalactic Lyman line blanketing following
\citet{Mad95}.
The error bars represent the $1\sigma$ uncertainties on the flux
measurements while upper limits indicate the $2\sigma$ confidence levels.
The solid lines show the template spectra for galaxy types E, Sbc, Scd,
and Im from \citet{Col80}, extended beyond the original wavelength range
using evolutionary synthesis models.  The shaded area outlines the range
in SED shapes covered by these templates within the original wavelength
limits.  In both panels, all SEDs have been normalized to unit flux at
$\rm \lambda_{\rm rest} = 2500$~\AA\ using linear interpolation between
the adjacent data points (the fluxes are in units of
$\rm erg\,s^{-1}\,cm^{-2}$\,\AA$^{-1}$).
The $J_{\rm s} - K_{\rm s} \geq 2.3$ selected objects have significantly
redder rest-frame optical colours than the LBGs, hence we refer to them
as ``distant red galaxies'' or DRGs.
\label{fig-rfSEDs}
}
\end{figure}

\begin{figure}[p]
\epsscale{0.80}
\plotone{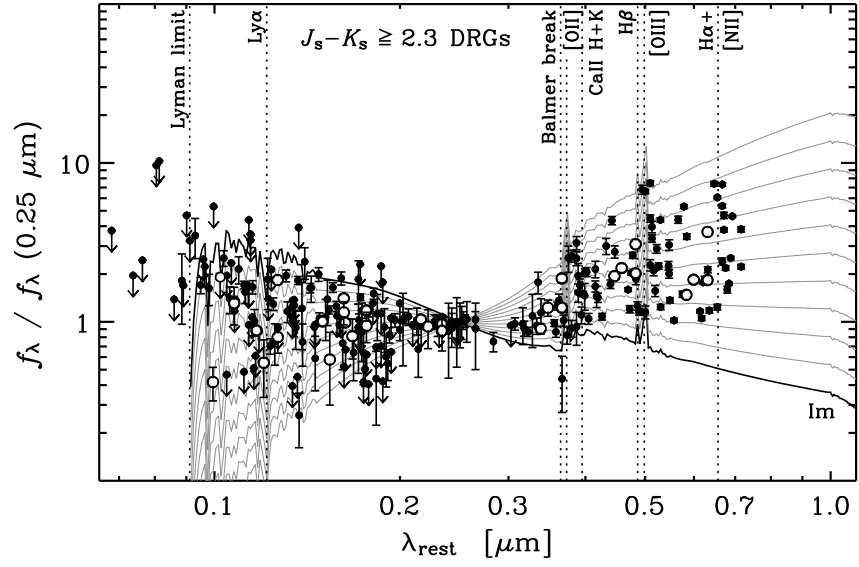}
\vspace{0.0cm}
\caption{
Same as Figure~\ref{fig-rfSEDs} for the rest-frame SEDs
of $2 \leq z \leq 3.5$ DRGs in the FIRES fields.
Only the Im galaxy template from \citet{Col80} is shown here
({\em black line\/}) and used to illustrate the effects of interstellar
extinction ({\em grey lines\/}) for $A_{V}$ between 0 and 3~mag, increasing
in steps of 0.3~mag.  The Im template itself includes intrinsic reddening
so that our simulations represent additional amounts of extinction.
The extinction was applied using the \citet{Cal00} law.
\label{fig-rfSEDs_sim}
}
\end{figure}

\begin{figure}[p]
\epsscale{0.9}
\plotone{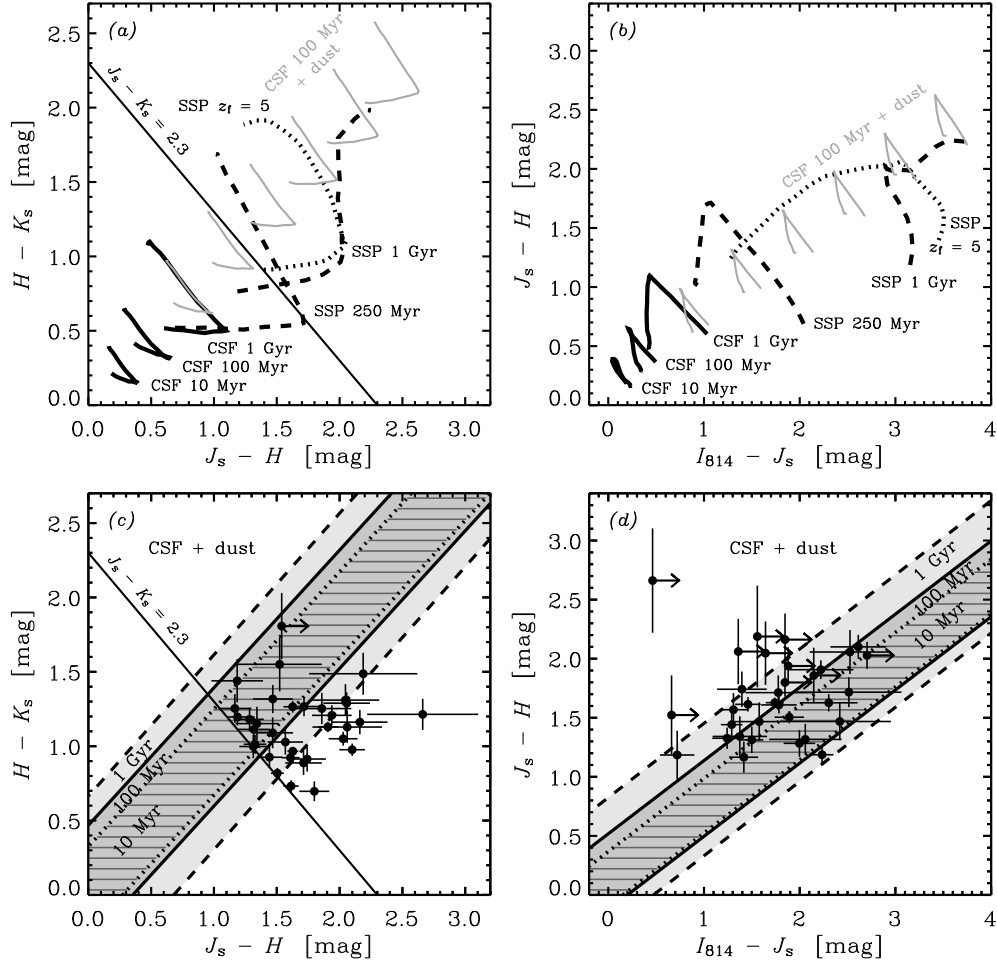}
\vspace{0.5cm}
\caption{
Diagnostic $I_{814}\,J_{\rm s}\,H\,K_{\rm s}$ colour diagrams 
to distinguish between dust obscuration and age effects.
({\em a\/}) $H - K_{\rm s}$ versus $J_{\rm s} - H$.  The diagonal
line indicates the $J_{\rm s} - K_{\rm s} = 2.3~{\rm mag}$ colour
cutoff used to select DRGs.
The various tracks show the evolution with redshift in the interval
$2 \leq z \leq 4$ of the observed colours of synthetic stellar populations.
The template SEDs were generated with the \citet{Bru03} synthesis code,
assuming a Salpeter IMF between 0.1 and $\rm 100~M_{\odot}$ and solar
metallicity.  The solid black lines correspond to models with constant
star formation rate (CSF) and fixed ages of 10~Myr, 100~Myr, and 1~Gyr.
The black dashed lines represent single stellar populations (SSPs) with
fixed ages of 250~Myr and 1~Gyr.  The black dotted line is the track
for an SSP formed at $z_{\rm f} = 5$ and passively evolving with time.
The grey solid lines show the effects of extinction on the 100~Myr old
CSF model, using the \citet{Cal00} extinction law; the $A_{V}$ increases
from 1 to 6~mag (in steps of 1~mag) for the tracks from the bottom left
to the top right.
({\em b\/}) Same as ({\em a\/}) for
$J_{\rm s} - H$ versus $I_{814} - J_{\rm s}$.
({\em c\/}) Colour distribution in the
$H - K_{\rm s}$ versus $J_{\rm s} - H$ diagram of the combined sample of
$2 \leq z \leq 3.5$ DRGs in the HDF-S and \mscl\ fields ({\em black dots\/}).
The error bars represent the $1\sigma$ measurements uncertainties and limits
are plotted at the $2\sigma$ confidence levels.  The shaded areas indicate
the regions occupied by the tracks for dusty CSF models with fixed ages of
10~Myr ({\em hatched region between dotted lines\/}),
100~Myr ({\em dark grey shaded region between solid lines\/}),
and 1~Gyr ({\em light grey shaded region between dashed lines\/}).
({\em d\/}) Same as ({\em c\/}) in the
$J_{\rm s} - H$ versus $I_{814} - J_{\rm s}$ diagram.
In both diagrams, the spread of colours of DRGs in the direction perpendicular
to the extinction paths is inconsistent with very young ($\rm \la 100~Myr$)
populations highly obscured by dust; the presence of more evolved
($\rm \ga 250~Myr$) populations with prominent Balmer/4000\,\AA\ break
is required for a large fraction of DRGs.
\label{fig-NIRcol}
}
\end{figure}

\begin{figure}[p]
\epsscale{0.9}
\plotone{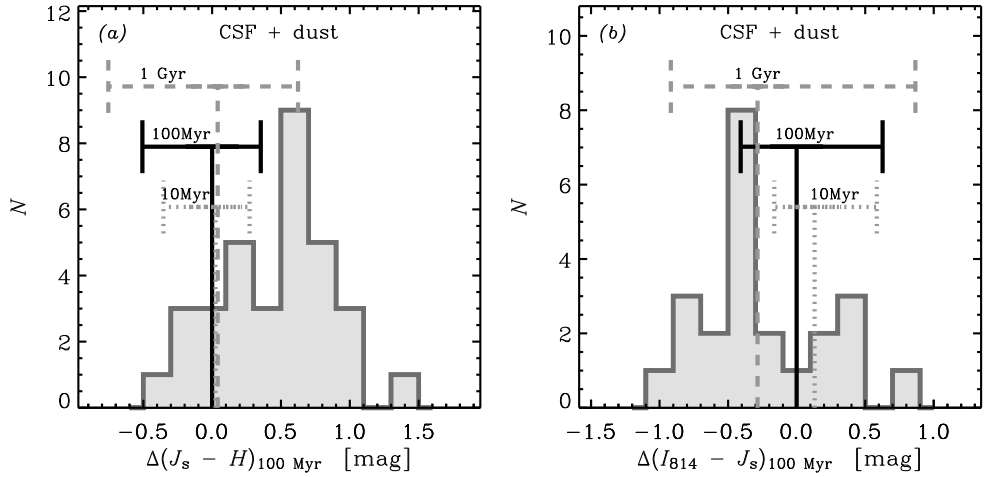}
\vspace{0.5cm}
\caption{
Colour excesses relative to dusty constant star formation models
of the combined sample of $2 \leq z \leq 3.5$ DRGs in the HDF-S
and $\rm MS\,1054-03$ fields.
({\em a\/}) Histogram of the deviation in $J_{\rm s} - H$ colour of
DRGs relative to the central locus of the dust-attenuated 100~Myr old
CSF model tracks for $2 \leq z \leq 4$ in the $H - K_{\rm s}$ versus
$J_{\rm s} - H$ diagram
(based on Figure~\ref{fig-NIRcol}, panels {\em a\/} and {\em c\/}).
The one object with a limit on its $J_{\rm s} - H$ colour is excluded.
The black solid horizontal error bar indicates the maximum range in
$J_{\rm s} - H$ colour of the models around the central locus shown
by the vertical black solid line at $\Delta(J_{\rm s} - H) = 0$.
Similarly, the dotted and dashed grey error bars indicate the ranges
for dusty CSF models of ages 10~Myr and 1~Gyr around their respective
central locus, plotted with the corresponding vertical lines at their
mean $J_{\rm s} - H$ deviation from that of the 100~Myr old models.
({\em b\/}) Same as ({\em a\/}) for the deviation in $I_{814} - J_{\rm s}$
colour in the $J_{\rm s} - H$ versus $I_{814} - J_{\rm s}$ diagram
(based on Figure~\ref{fig-NIRcol}, panels {\em b\/} and {\em d\/}).
The reference set of models is again the suite of dusty 100~Myr old CSF
populations, with central locus at $\Delta(I_{814} - J_{\rm s}) = 0$.
The ten objects with limits on their $I_{814} - J_{\rm s}$ or
$J_{\rm s} - H$ colour are excluded.
\label{fig-NIRcol_proj}
}
\end{figure}

\begin{figure}[p]
\epsscale{0.90}
\plotone{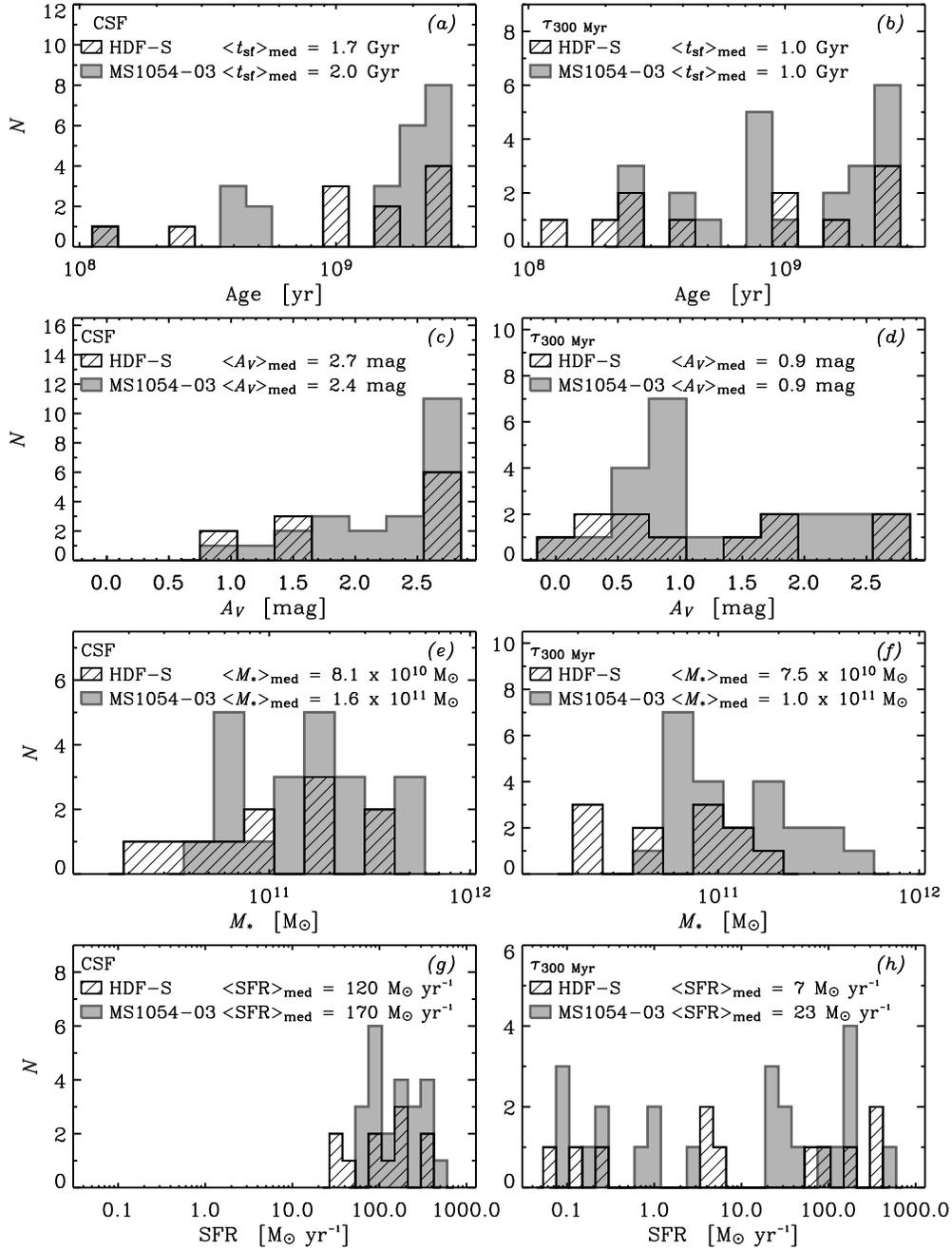}
\vspace{0.5cm}
\caption{
Distributions of best-fit properties of the DRG samples in the FIRES fields.
The panels to the left show the model results for a constant star formation
rate (CSF) and to the right, for an exponentially declining star formation
rate with $e$-folding timescale $\rm \tau = 300\,Myr$ ($\tau_{\rm 300\,Myr}$).
The models assumed a Salpeter IMF between 0.1 and $\rm 100~M_{\odot}$,
solar metallicity, and the \citet{Cal00} extinction law.
The results are shown separately for the sample in HDF-S
({\em hatched histograms\/}) and in the $\rm MS\,1054-03$ field
({\em grey shaded histograms\/}), and the median values are given
in each panel.
({\em a\/}) and ({\em b\/}) Best-fit ages $t_{\rm sf}$.
({\em c\/}) and ({\em d\/}) Best-fit extinctions $A_{V}$.
({\em e\/}) and ({\em f\/}) Stellar masses $M_{\star}$ derived
from the best-fitting synthetic models.
({\em g\/}) and ({\em h\/}) Same as ({\em e\/}) and ({\em f\/})
for the instantaneous star formation rates (SFRs).
\label{fig-res_std2}
}
\end{figure}

Both the CSF and \tauint\ models provide acceptable fits to the SEDs
of our DRG samples
\footnote{The median of the chi-squared values normalized per degree of
freedom, $\chi^{2}_{\rm n}$, for the best-fit CSF models are 1.6 for the
HDF-S sample and 2.1 for the \mscl\ field sample.  For the \tauint\ models,
the $\chi^{2}_{\rm n}$ values are 0.6 and 1.6, respectively.
$\chi^{2}_{\rm n}$'s higher than the expected value of unity for good fits
could indicate that the models do not provide a realistic representation of
the objects, the photometric uncertainties underestimate the real measurement
errors, the distributions of uncertainties of the modeled properties are not
Gaussian, or, most likely, that a combination of the factors is at play.}.
The best-fitting models for selected sources are shown in Figure~\ref{fig-SEDs}.
Table~\ref{tab-res_std2} and Figure~\ref{fig-res_std2} summarize the derived
properties.  The results are similar for the HDF-S and \mscl\ field samples.
With the CSF model, the median values for HDF-S (\mscl\ field) are
$\langle t_{\rm sf} \rangle_{\rm med} = 1.7$ (2.0)~Gyr,
$\langle A_{V} \rangle_{\rm med} = 2.7$ (2.4)~mag,
$\langle M_{\star} \rangle_{\rm med} = 
  8.1~(16) \times 10^{10}~{\rm M_{\odot}}$,
$\langle M_{\star} / L_{V, \star} \rangle_{\rm med} =
  1.2~(2.3)~{\rm M_{\odot}\,L_{V,\odot}^{-1}}$, and instantaneous
$\langle {\rm SFR} \rangle_{\rm med} = 120~(170)~{\rm M_{\odot}\,yr^{-1}}$.
With the \tauint\ model, the median age and extinction are 1.0~Gyr and
0.9~mag for both fields; the median for the other properties in HDF-S
(\mscl\ field) are $7.5~(10) \times 10^{10}~{\rm M_{\odot}}$,
$1.2~(1.8)~{\rm M_{\odot}\,L_{V,\odot}^{-1}}$, and 
$7~(23)~{\rm M_{\odot}\,yr^{-1}}$.
The median ages of the bulk of stars $\langle t_{\rm sf} \rangle_{\rm SFR}$
and the median time-averaged SFRs $M_{\star} / t_{\rm sf}$ for the two
SFHs and two fields differ less than the median $t_{\rm sf}$ and
instantaneous SFRs, with values between 0.75 and 1.0~Gyr, and 90
and $140~{\rm M_{\odot}\,yr^{-1}}$.

The dispersion among individual DRGs is large for all properties.  This
stems partly from the diversity in SEDs noted in \S~\ref{Sect-prop}.
For instance, with a given SFH, quite different combinations of age and
extinction are derived for the objects that are bluest in the rest-frame
UV compared to the reddest ones.  Part of the (small) differences between
the HDF-S and \mscl\ samples likely reflects the brighter $K_{\rm s}$-band
magnitude limit and larger proportion of the brightest sources, the redder
$J_{\rm s} - K_{\rm s}$ colours, and the lower redshifts for the \mscl\
objects (see Table~\ref{tab-sample}).
The different sets of optical photometric bands may play a role
too by providing somewhat different observational constraints.

For the ensemble of DRGs, either SFH thus implies old ages, and high
stellar masses and M/L ratios.  High SFRs and large amounts of extinction
are derived for the CSF model while the \tauint\ model leads to lower SFRs
and extinction (we comment on the validity of the adopted \citealt{Cal00}
extinction law in view of the large extinction derived for the DRGs in
\S~\ref{Sub-paramvar_AVlaw_Z}).
Similar results were obtained in our initial study of HDF-S
\citep{Fra03} and for the spectroscopic sample \citep{Dok04}.  In particular,
for the two spectroscopically-confirmed non-AGN DRGs in the \mscl\ field, the
ages are supported by estimates from the H$\alpha$ equivalent widths and the
stellar masses are consistent with the dynamical masses from the H$\alpha$
or [\ion{O}{3}]\,$\lambda 4959, 5007$~\AA\ line widths.
As we show in \S~\ref{Sub-paramvar}, the stellar masses are the
most robust properties against variations in SFH, metallicity, and
extinction law (neglecting changes in the IMF).  With our definition,
the $M_{\star} / L_{V, \star}$ ratios also are because the luminosities
involved are extrinsic.  On the other hand, the instantaneous SFRs are
very sensitive to the model assumptions, especially to the SFH (again
ignoring the IMF).

To test the sensitivity to the photometric redshift uncertainties, we modeled
the DRG samples as above, adopting in turn the lowest and highest redshift
allowed by the 68\% confidence interval around $z_{\rm ph}$ for each object.
Although the effects on the median values of the best-fit properties can be
appreciable (by up to 0.7~Gyr for the age, 0.9~mag for $A_{V}$, a factor of 2
for $M_{\star}$, and a factor of 5 for SFR, depending on the field and SFH),
our conclusions are qualitatively unaffected.
We also modeled the samples of DRGs including the objects outside of the
$z = 2 - 3.5$ range.  The impact on the median properties is significantly
smaller ($\rm \leq 0.5~Gyr$, $\rm \leq 0.3~mag$, $< 15\%$, and $< 30\%$
for the ages, $A_{V}$'s, $M_{\star}$'s, and SFRs, respectively).

The SED fits of Figure~\ref{fig-SEDs} illustrate well the difficulty
in discriminating between the two SFHs (in particular, e.g., for
$\rm HDFS-496$ and $\rm MS-1383$).  
Based on the formal $\chi^2$ values, the fits are statistically
indistinguishable for all but three galaxies, for which the CSF model could
in principle be ruled out at the 95\% confidence level.  For two of these,
$\rm HDFS-767$ and $\rm MS-1038$, the higher flux in the $H$ band relative to
the $K_{\rm s}$ band strongly influences this result.  Although their
$H - K_{\rm s}$ colour is fairly blue, they are not the most extreme of our
sample.  For the other, $\rm MS-1319$, the optical data are better reproduced
with the \tauint\ model.  For this spectroscopically studied DRG \citep{Dok04},
correcting for line emission would affect mostly the observed $B$-band flux
(downwards by 10\% for Ly$\alpha$) and would improve the agreement with the
CSF model.  The distinction between the models is not very robust and the
$\chi^2$ computation relies on uncertainties that do not account for those
of $z_{\rm ph}$ or possible emission line contributions.  In view of this,
we chose not to discriminate between SFH models for individual sources in
our analysis.

\subsection{Variations in Model Parameters}   \label{Sub-paramvar}

\begin{figure}[p]
\epsscale{0.90}
\plotone{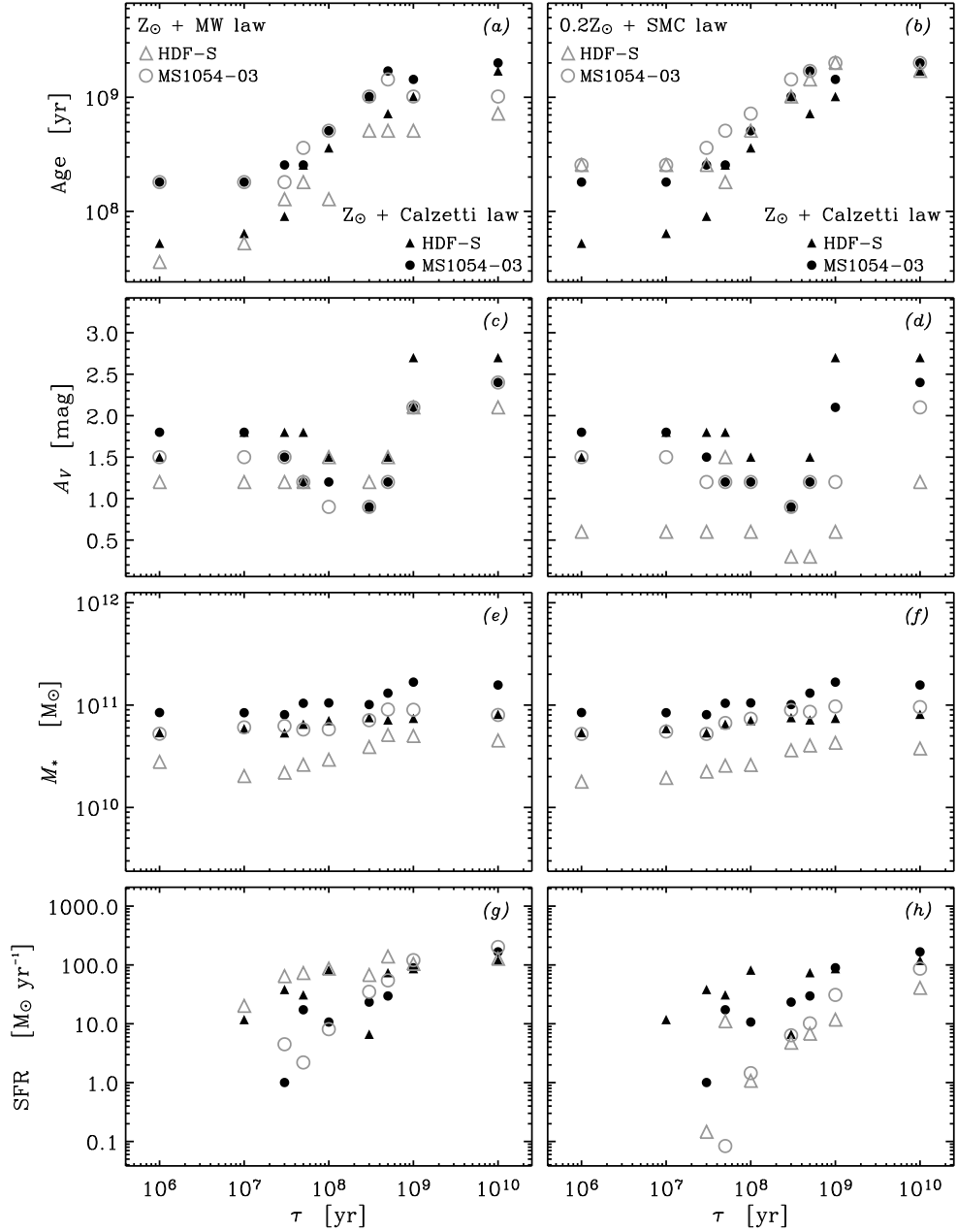}
\vspace{0.5cm}
\caption{
Variations of the best-fit properties of DRGs with assumptions
on the star formation history, extinction law, and metallicity.
The median values for the samples in the HDF-S ({\em triangles\/})
and the $\rm MS\,1054-03$ field ({\em circles\/}) are plotted
for a suite of exponentially declining SFRs with $e$-folding
timescale $\tau$ between 10~Myr and 1~Gyr.  Median values are
also shown for single stellar population models 
(SSP; effective $\tau = 0$, plotted at $\rm \tau = 10^{6}~yr$)
and for models with constant star formation rate
(CSF; effective $\tau = \infty$, plotted at $\rm \tau = 10^{10}~yr$).
The filled symbols in all panels represent results with solar metallicity
$Z = {\rm Z_{\odot}}$ and the \citet{Cal00} extinction law.  The open
symbols in the panels to the left show the results with the Milky Way
extinction law \citep[MW;][]{All76} and keeping $Z = {\rm Z_{\odot}}$.
In the panels to the right, open symbols show the results with
$Z = 0.2~{\rm Z_{\odot}}$ and the Small Magellanic Cloud extinction
law \citep[SMC;][]{Pre84, Bou85}.  A Salpeter IMF between 0.1 and
$\rm 100~M_{\odot}$ was assumed in all cases.
({\em a\/}) and ({\em b\/}) Best-fit ages $t_{\rm sf}$.
({\em c\/}) and ({\em d\/}) Best-fit extinctions $A_{V}$.
({\em e\/}) and ({\em f\/}) Best-fit stellar masses $M_{\star}$.
({\em g\/}) and ({\em h\/}) Best-fit instantaneous star formation rates (SFRs).
The SFR for the SSP models are $\rm 0~M_{\odot}\,yr^{-1}$.
Other values outside of the plotted range
are as follows (in units of $\rm M_{\odot}\,yr^{-1}$).
$\rm MS\,1054-03$ for $Z = {\rm Z_{\odot}}$, \citeauthor{Cal00} law, and
$\rm \tau = 10^{7}~yr$: $\rm 9.8 \times 10^{-4}$.
$\rm MS\,1054-03$ for $Z = {\rm Z_{\odot}}$, MW law, and
$\rm \tau = 10^{7}~yr$: $\rm 6.0 \times 10^{-4}$.
$\rm MS\,1054-03$ for $Z = 0.2\,{\rm Z_{\odot}}$ and SMC law:
$\rm 2.7 \times 10^{-8}$ for $\rm \tau = 10^{7}~yr$ and
$\rm 8.2 \times 10^{-3}$ for $\rm \tau = 3 \times 10^{7}~yr$.
HDF-S for $Z = 0.2\,{\rm Z_{\odot}}$, SMC law, and
$\rm \tau = 10^{7}~yr$: $\rm 1.3 \times 10^{-8}$.
\label{fig-paramvar}
}
\end{figure}

We explored the effects of variations in three of the model assumptions:
the SFH, the extinction law, and the metallicity.  Figure~\ref{fig-paramvar}
summarizes the results for the ages, extinction, stellar masses, and SFRs
derived for different sets of input parameters.  While the effects may differ
among individual objects, we are here interested in the ensemble properties
of DRGs and so the discussion refers to the median values.  We find that the
SFH has the largest impact on the derived quantities.  The age, extinction,
and instantaneous SFR are the most affected ones whereas the stellar mass
and M/L ratio are the least sensitive ones and change only by factors of
$\approx 2 - 3$.

\subsubsection{Star Formation History}   \label{Sub-paramvar_SFH}

We considered exponentially declining SFRs with a range of $e$-folding
timescales $\tau$ between 10~Myr and 1~Gyr.  They are compared with the
CSF and also an SSP model, equivalent to $\tau = \infty$ and $\tau = 0$,
respectively.  With our broad-band data, it is not possible to discriminate
among any of the SFHs for the large majority of our DRGs.
The overall trend with decreasing timescale is of younger ages by up
to an order of magnitude.  The extinction values become lower but do not
vary monotonically with $\tau$, exhibiting a minimum around 300~Myr.
The maximum difference amounts to 1.8~mag, or a factor of $\approx 5$
in attenuation at $V$ in the rest-frame.
It is intriguing that the lowest extinction values are not derived for
the models with shortest timescales, as might be expected.  This may be
due to worse template mismatch at $\rm \tau < 10^{8}~yr$.
The SFRs drop by several orders of magnitude, to some extent because of
lower extinction but mainly as a consequence of the functional form of
the assumed SFR itself combined with the ages, which are such that
$R(t_{\rm sf}) / R_{0} = \exp(-t_{\rm sf}/\tau) \ll 1$.

The stellar masses tend to decrease with smaller $\tau$ but remain within a
factor of 2 of the masses for CSF.  We thus reach similar conclusions
about the stability of derived $M_{\star}$ for DRGs as \citet{Pap01}
and \citet{Sha01} in their modeling of LBGs also based on broad-band
optical-to-NIR photometry.  As discussed by these authors, this is in part
due to the NIR data probing the rest-frame optical regime, which is less
sensitive to $M/L$ variations due to age or extinction than the rest-frame
UV.  In addition, the age and dust degeneracies tend to cancel out in terms
of the effect on the ratio of stellar mass to observed (attenuated) light.
The same behaviour is seen in the case of our DRGs
\citep[see][for a discussion]{Dok04}.
While age and extinction can vary strongly with the SFH, the combinations
of best-fit values result in much tighter constraints on $M_{\star}$ and
$M_{\star}/L_{V,\star}$.

The largest extinction is derived for the longest timescales.  The need
for high extinction here may be artificially enhanced by the choice of
SFH along with the constraint that the best-fit ages do not exceed that
of the universe at each object's redshift.  In the limit of CSF, the
population of young massive stars rapidly levels off on the timescale of
their main-sequence lifetime ($\rm \sim 10^{6} - 10^{7}~yr$) while the
population of lower-mass longer-lived stars grows progressively.  To match
the typically red SEDs of DRGs, the UV light from the luminous massive
stars continuously formed has to be suppressed with sufficient extinction.
The maximum age constraint limits the relative contribution of evolved
versus young stars to the integrated light, further enhancing the need
for extinction.
Over the range of SFHs explored here, the strong coupling between age and
$\tau$ also reflects the need for the build-up of a large enough population
of evolved stars.

Models with very short timescales $\rm < 10^{8}~yr$ where the extinction
is allowed to vary lead to ages $\rm < 300~Myr$ and fairly high 
$A_{V} \approx 1.5~{\rm mag}$.  Arguably, such young dusty burst models may
not be appropriate for the ensemble of DRGs (though they may be for some
individual objects) in view of the evidence for a prominent spectral break
in their rest-frame continuum SEDs (see \S~\ref{Sect-prop}).  Choices 
of $\rm \tau \sim 10^{8}~yr$ lead to ages consistent with the appearance of
a strong Balmer/4000\,\AA\ break in both cases of very quiescent or still
actively star-forming galaxies.  Although models with $\rm \tau \la 10^{8}~yr$
cannot be statistically ruled out in our fits, we regard them as least
plausible for DRGs.

\subsubsection{Extinction Law and Metallicity}   \label{Sub-paramvar_AVlaw_Z}

To test the effects of possible variations in the extinction law and in
metallicity, we fitted two alternative suites of models to the SEDs.
In one variant,
we modified the prescription for the extinction law to that of \citet{All76}
for the Milky Way (MW), keeping the solar metallicity.  In the other, we used
the set of BC03 models for metallicity $Z = 0.2~{\rm Z_{\odot}}$ and adopted
for consistency the extinction law for the Small Magellanic Cloud 
\citep[SMC;][]{Pre84, Bou85}.
The main differences between the \citet{Cal00} and MW
extinction laws lie in the ratio of total-to-selective absorption
$R_{V} = A_{V} / E(B - V)$ (4.05 versus 3.1, respectively) and in the
\citeauthor{Cal00} law lacking the 2175~\AA\ bump characteristic of MW
dust mixtures.  Otherwise, their wavelength dependence are fairly similar.
The SMC law with $R_{V} = 2.72$ also lacks the 2175~\AA\ bump.  In addition,
it rises more steeply with decreasing wavelengths in the near-UV than the
other two laws; in other words, the \citeauthor{Cal00} and MW laws are much 
``greyer'' at near-UV wavelengths.

For the models with $Z = {\rm Z_{\odot}}$ and the MW law, the ages for a
given SFH remain the same or become mostly younger, by up to a factor of 3.
The best-fit $A_{V}$'s also remain the same or decrease, by at most 0.6~mag.
The effects are more important for the $Z = 0.2~{\rm Z_{\odot}}$ and SMC
law models, with generally older ages by up to a factor of 5 and $A_{V}$'s
lower by up to 2~mag.  In both model sets, the instantaneous SFRs are
strongly affected towards shorter timescales $\tau$, especially for the
low-metallicity case.  The stellar masses are systematically lower for
both variants, by a factor of $\approx 2$ on average.

The near-systematic increase in median age of the DRGs for the lower
metallicity models can be attributed to the combined effects from the
extinction law and stellar emission.  Since the SMC law is steeper in
the rest-frame UV than both the \citet{Cal00} and MW laws, lower $A_{V}$
values will reproduce the observed SED slopes in this regime but also imply
bluer rest-frame UV-to-optical colours such that older ages are required to
match the colours of DRGs.  In addition, at $Z < {\rm Z_{\odot}}$ and
for a fixed age, the stellar evolutionary tracks and model atmospheres
predict higher effective temperatures, bluer continua, and shallower
metallic and molecular absorption features, further driving up the
best-fit model ages.  As emphasized by \citet{Sha01},
it is important here to keep in mind that theoretical stellar tracks and
stellar atmosphere models for non-solar metallicities are still not fully
tested against empirical data \citep[see the discussion by][]{Bru03}.
The variations in derived parameters we describe above provide
nevertheless indications of the magnitude of metallicity effects.

We further note that the high obscuration inferred for the models with
star formation timescales $\rm \tau \ge 10^{9}~yr$ raises concerns about
the applicability to DRGs of the adopted \citet{Cal00} extinction law and
foreground dust screen geometry.  The \citeauthor{Cal00} law provides a
good representation for local UV-bright starbursts with $A_{V}$ up to about
2.5~mag and where the foreground dust approximation is observationally
supported.  For the CSF model, for instance, the $A_{V}$ values obtained for
half of the DRGs indicate extinction levels at the limit of, or beyond the
range of validity.  This may suggest that the \citeauthor{Cal00} law is
inappropriate, and that the dust and sources geometry is more complex.

\section{COMPARISON OF DRGS WITH LBGS AND OTHER $2 \leq z \leq 3.5$ GALAXIES}
         \label{Sect-discussion}

The picture that emerges from our SED analysis and modeling is that
DRGs constitute an old and massive galaxy population at high-redshift.
Obscuration by interstellar dust plays an important role in their
observed properties.  Depending on their actual SFH, they may also
still be very actively forming stars.  In this section, we interpret
our results in a broader context by comparing our DRGs with LBGs.
We also consider the ensemble of all $K_{\rm s}$-band selected galaxies
in both FIRES fields at $2 \leq z \leq 3.5$.
We modeled these samples in the exact same way as the DRGs.

\subsection{How Different are DRGs and LBGs?}
           \label{Sub-DRGs_LBGs}

An immediate implication of the model results presented in
\S~\ref{Sect-models} is that DRGs appear to be older, more obscured, and
with higher stellar masses than LBGs.  Our sample of optically-selected 
LBGs at $K_{\rm s}^{\rm tot} < 22.5~{\rm mag}$ and $2 \leq z \leq 3.5$ in
HDF-S allows a consistent comparison, based on the same data and with
identical model ingredients and assumptions.
With our standard CSF model, we derive median values for the LBGs of
$\langle t_{\rm sf} \rangle_{\rm med} = 500~{\rm Myr}$, 
$\langle A_{V} \rangle_{\rm med} = 0.6~{\rm mag}$,
$\langle M_{\star} \rangle_{\rm med} = 1.4 \times 10^{10}~{\rm M_{\odot}}$,
$\langle M_{\star} / L_{V, \star} \rangle_{\rm med} =
  0.4~{\rm M_{\odot}\,L_{V,\odot}^{-1}}$,
and $\langle {\rm SFR} \rangle_{\rm med} = 39~{\rm M_{\odot}\,yr^{-1}}$.
With the \tauint\ model, we find lower median age of 360~Myr and SFR of
$21~{\rm M_{\odot}\,yr^{-1}}$, but the same median extinction, stellar
mass, and M/L ratio.  We cannot statistically discriminate between the
fits for different SFHs for any of our LBGs.

Compared to the HDF-S LBGs and with CSF, our DRGs are typically $3 - 4$
times older, $4 - 5$ times more obscured, have roughly an order of magnitude
higher stellar mass, and form stars at $3 - 5$ times higher instantaneous
rates (based on the median properties).  Among the various declining SFHs
we considered, the \tauint\ model leads to the lowest extinction for the DRGs
(Figure~\ref{fig-paramvar}).  One might expect that this SFH reduces most the
inferred differences between the two populations.  The DRGs, however, remain
distinct from LBGs in their median properties except for the SFR, being 3
times older, 1.5 times more extincted, $5 - 8$ times more massive, and
$0.3 - 1.3$ times as actively star-forming.
As for the DRGs, the LBGs show an appreciable dispersion in best-fit properties
with some overlap with DRGs.  However, the differences in the median values and
combinations thereof (especially age, $A_{V}$, and $M_{\star}$) indicate that
the ensemble properties of the two populations are different for either SFH.

\begin{figure}[p]
\epsscale{0.65}
\plotone{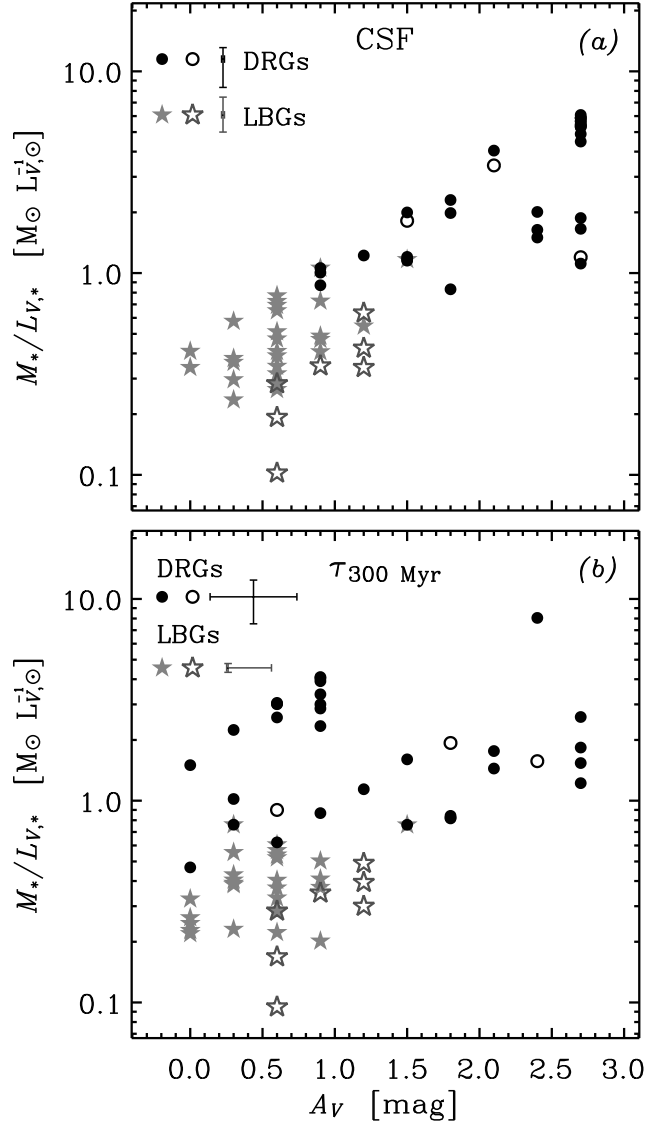}
\vspace{0.5cm}
\caption{
Distribution in the $M_{\star}/L_{V,\star} - A_{V}$ plane of
the $2 \leq z \leq 3.5$ DRG and LBG samples in the FIRES fields.
The data are plotted for the combined DRG samples in HDF-S at
$K_{\rm s}^{\rm tot} < 22.5~{\rm mag}$ and in the $\rm MS\,1054-03$
field at $K_{\rm s}^{\rm tot} < 21.7~{\rm mag}$ ({\em circles\/}),
and for the optically-selected LBGs in HDF-S at
$K_{\rm s}^{\rm tot} < 22.5~{\rm mag}$ ({\em stars\/}).
Objects with and without spectroscopic redshift are distinguished
({\em open and filled symbols\/}, respectively).
The error bars indicate the median of the 68\% confidence intervals derived
individually for each object in the DRG and the LBG samples as explained in
\S~\ref{Sub-model_proc} 
(because of our sampling of the extinction parameter space with intervals
$\Delta A_{V} = 0.3~{\rm mag}$, median values of the confidence intervals
for $A_{V}$ that are formally equal to 0~mag could actually be larger but
$< 0.3~{\rm mag}$). 
The $L_{V,\,\star}$ is the stellar rest-frame attenuated $V$-band luminosity.
({\em a\/}) Results for the model with constant star formation rate (CSF).
({\em b\/}) Same as ({\em a\/}) for the exponentially declining
star formation rate model with $e$-folding timescale
$\rm \tau = 300~Myr$ ($\tau_{\rm 300\,Myr}$).
All models assumed a Salpeter IMF between 0.1 and $\rm 100~M_{\odot}$,
solar metallicity, and the \citet{Cal00} extinction law.
\label{fig-MLV_AV}
}
\end{figure}

The distinction between the DRGs and LBGs is best illustrated in
Figure~\ref{fig-MLV_AV}, which compares the distribution of the
two populations in the $M_{\star}/L_{V,\star} - A_{V}$ plane
(where the median uncertainties from the 68\% confidence intervals are
$\la 25\%$ in $M_{\star}/L_{V, \star}$ and $\leq 0.3~{\rm mag}$ in $A_{V}$
for the DRG and LBG samples, and our two standard models).
In this parameter space, the overlap is very small for the CSF model.
For the \tauint\ model, about two-thirds of the DRGs shift to a lower
extinction interval $A_{V} < 1.5~{\rm mag}$, in common with the LBGs,
while the rest uniformly populates the higher $A_{V}$ range explored.
The ensemble of DRGs and LBGs remain however clearly separated in the
ranges of $M_{\star}/L_{V,\star}$ ratios covered, with very little
overlap.  This is driven by the stellar masses since the DRG and LBG
samples are similarly luminous in the rest-frame $V$ band, with nearly
identical median values of $\rm 5 \times 10^{10}~L_{V,\odot}$ (within 20\%).

The weakness of our HDF-S LBG sample is that $z_{\rm sp}$'s are available 
for seven out of 33 galaxies.  However, we arrive at similar conclusions if
we consider the model results for the spectroscopically confirmed LBG samples
of \citet{Pap01} and \citet{Sha01}, also based on optical to NIR SEDs.
The typical (median or geometric mean) ages inferred are
$\rm \sim 10^{8}~yr$, with extinction $A_{V} < 1~{\rm mag}$ and
$M_{\star} \sim 10^{10}~{\rm M_{\odot}}$ (for $Z = {\rm Z_{\odot}}$,
a Salpeter IMF down to 0.1\masssol, and our adopted cosmology).
Our results are within a factor of a few of those of \citeauthor{Pap01}
and \citeauthor{Sha01}, similar to the difference between the latter two.
It is however difficult to compare in detail the results among all three
studies.  The data sets are different in terms of depth, quality,
and photometric bands involved.
The sample of \citet{Pap01} was selected in the HDF-N from the space-based
HST data while the ground-based sample of \citet{Sha01} was selected over a
much wider area and is brighter.  In their modeling, \citet{Pap01} assumed
exponentially declining SFHs with the timescale as a free parameter.
\citet{Sha01} did not attempt to fit the SFH and discuss mainly the
CSF scenario.  In addition, the models of \citeauthor{Pap01} and
\citeauthor{Sha01} relied on earlier versions of the Bruzual \& Charlot
code and partly different model ingredients.  In view of this, it is
probably fair to say that the properties derived for all three LBG
samples are in broad agreement.  The conclusions about the differences
between DRGs and LBGs remain qualitatively the same.

\subsection{Derived Properties versus Observed NIR Properties}
           \label{Sub-modprop_NIRprop}

We investigated relationships between the modeled quantities and the
observed NIR properties, the key feature of our DRG selection.  For this
purpose, we complemented the DRG samples with the HDF-S LBGs and all other
$K_{\rm s}$-band selected galaxies at $2 \leq z \leq 3.5$ in both the HDF-S
and \mscl\ fields down to the same $K_{\rm s}^{\rm tot}$ limits of
22.5 and 21.7~mag, respectively.

\begin{figure}[p]
\epsscale{0.95}
\plotone{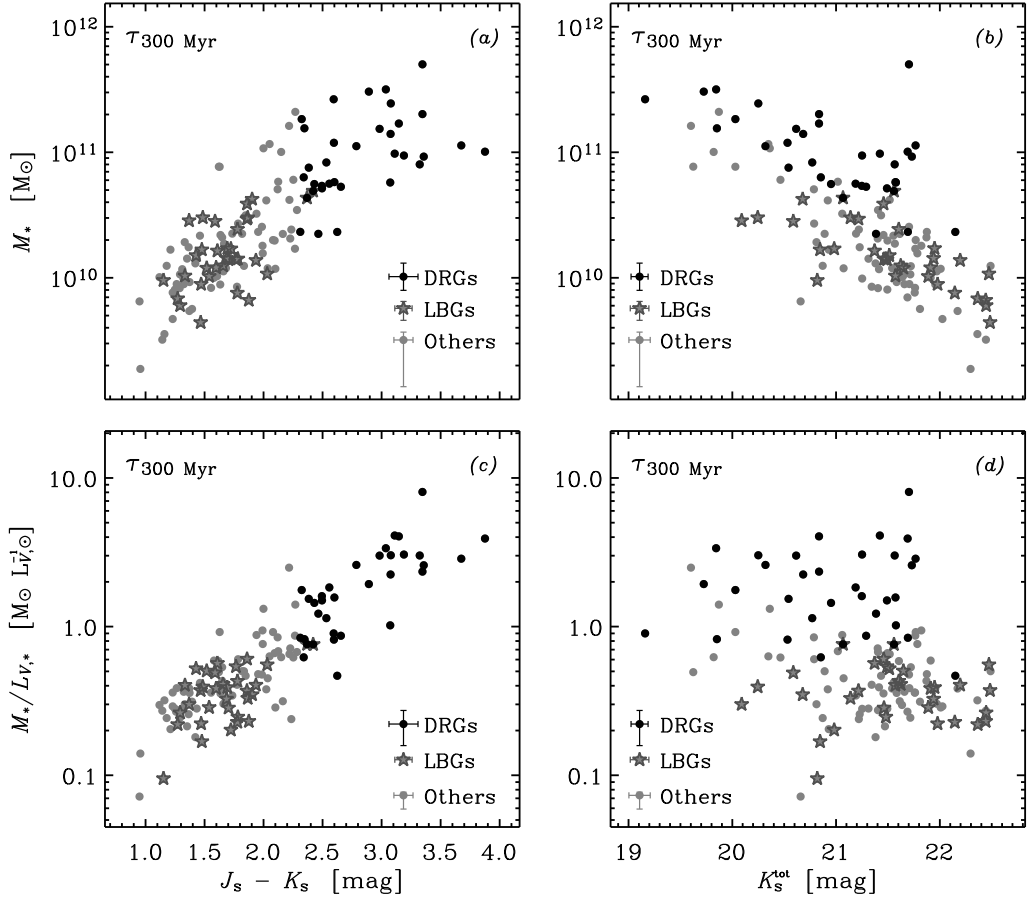}
\vspace{0.5cm}
\caption{
Variations of derived properties of $2 \leq z \leq 3.5$ galaxy populations
in the FIRES fields as a function of observed NIR properties.  The results
are shown for the exponentially declining $\tau_{\rm 300\,Myr}$ model, and
are qualitatively similar for the constant star formation rate model. 
Different symbols are used for the combined DRG samples in the HDF-S at
$K_{\rm s}^{\rm tot} < 22.5~{\rm mag}$ and the $\rm MS\,1054-03$ field
at $K_{\rm s}^{\rm tot} < 21.7~{\rm mag}$ ({\em black dots\/}), the
optically-selected LBGs in the HDF-S at $K_{\rm s}^{\rm tot} < 22.5~{\rm mag}$
({\em grey stars\/}), and all other $K_{\rm s}$-band selected sources
at $2 \leq z \leq 3.5$ in both fields and down to the respective 
$K_{\rm s}^{\rm tot}$ magnitude limits applied for DRGs and LBGs
({\em grey dots\/}).
The error bars indicate the median of the 68\% confidence intervals derived
individually for each object in the three samples (see \S~\ref{Sub-model_proc}).
({\em a\/}) Stellar mass $M_{\star}$ versus $J_{\rm s} - K_{\rm s}$ colour.
({\em b\/}) Stellar mass $M_{\star}$ versus $K_{\rm s}^{\rm tot}$ magnitude.
({\em c\/}) and ({\em d\/}) Same as ({\em a\/}) and ({\em b\/}) for the ratio
of stellar mass to rest-frame attenuated $V$-band luminosity
$M_{\star}/L_{V, \star}$.
The models assumed a Salpeter IMF between 0.1 and $\rm 100~{\rm M_{\odot}}$,
solar metallicity, and the \citet{Cal00} extinction law.
\label{fig-resall1}
}
\end{figure}

Figure~\ref{fig-resall1} plots, for our standard \tauint\ model,
the derived stellar masses and M/L ratios as a function of observed
$J_{\rm s} - K_{\rm s}$ colour and $K_{\rm s}$-band magnitude.
The results with the CSF are qualitatively similar.
There is a clear correlation between the derived $M_{\star}$ and the
observed $J_{\rm s} - K_{\rm s}$ colour, which is even tighter for the
$M_{\star} / L_{V,\,\star}$ ratio.  The redder the objects, the higher
the stellar masses and mass-to-light ratios.  The DRGs lie at the massive
end and extend smoothly the relationships for the $2 \leq z \leq 3.5$
objects with bluer $J_{\rm s} - K_{\rm s}$ colours, including the LBGs.
Because of the different limiting magnitudes between the HDF-S and \mscl\
fields, it is difficult to draw conclusions on the variations of $M_{\star}$ 
and $M_{\star}/L_{V,\,\star}$ versus $K_{\rm s}^{\rm tot}$.  The data suggest
that our $K_{\rm s}$-band selection does not miss a significant fraction of
massive galaxies at $2 \leq z \leq 3.5$ that would be faint because of
extinction effects.

\begin{figure}[p]
\epsscale{0.95}
\plotone{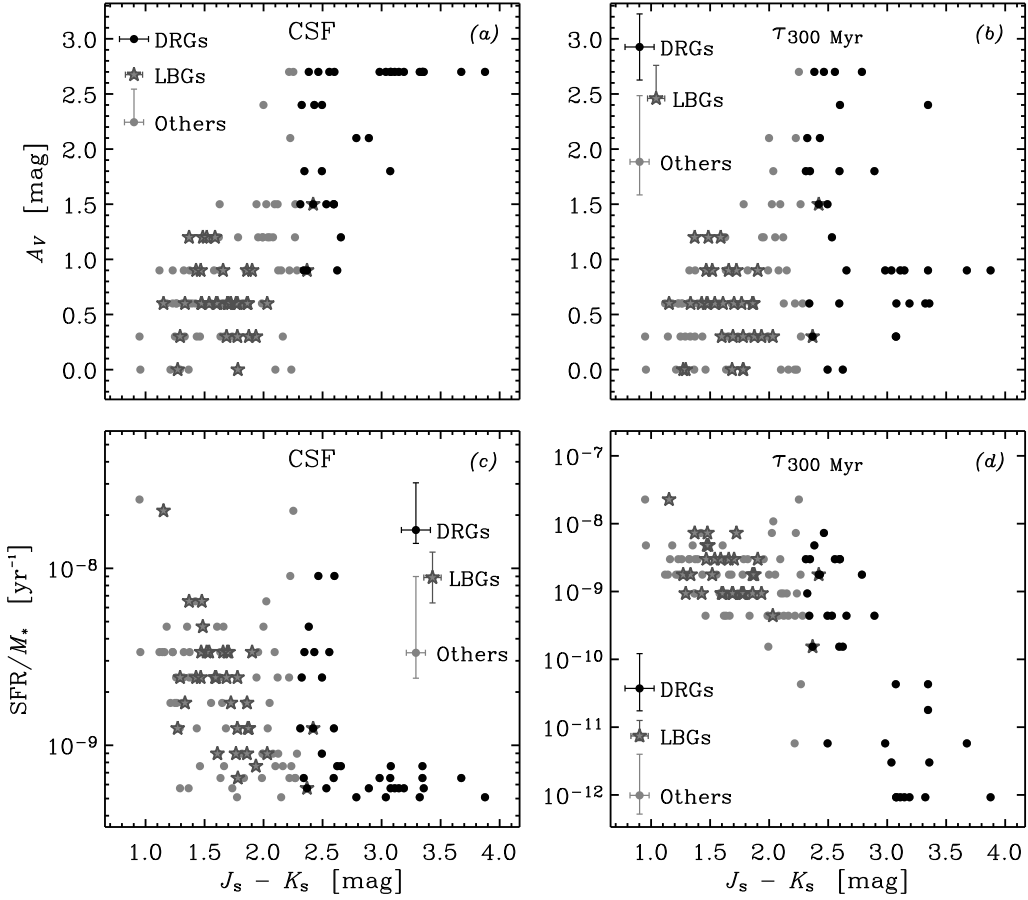}
\vspace{0.5cm}
\caption{
Variations of derived properties of $2 \leq z \leq 3.5$ galaxy populations
in the FIRES fields as a function of observed $J_{\rm s} - K_{\rm s}$ colour.
Different symbols are used for the combined DRG samples in the HDF-S at
$K_{\rm s}^{\rm tot} < 22.5~{\rm mag}$ and the $\rm MS\,1054-03$ field
at $K_{\rm s}^{\rm tot} < 21.7~{\rm mag}$ ({\em black dots\/}), the
optically-selected LBGs in the HDF-S at $K_{\rm s}^{\rm tot} < 22.5~{\rm mag}$
({\em grey stars\/}), and all other $K_{\rm s}$-band selected sources
at $2 \leq z \leq 3.5$ in both fields and down to the respective 
$K_{\rm s}^{\rm tot}$ magnitude limits applied for DRGs and LBGs
({\em grey dots\/}).
The error bars indicate the median of the 68\% confidence intervals derived
individually for each object in the three samples (see \S~\ref{Sub-model_proc};
because of our sampling of the extinction parameter space with intervals
$\Delta A_{V} = 0.3~{\rm mag}$, median values of the confidence intervals
for $A_{V}$ that are formally equal to 0~mag could actually be larger but
$< 0.3~{\rm mag}$).
({\em a\/}) Extinction $A_{V}$ for the model with constant star formation rate
(CSF).
({\em b\/}) Extinction $A_{V}$ for the model with exponentially declining
$\tau_{\rm 300\,Myr}$ model.
({\em c\/}) and ({\em d\/}) Same as ({\em a\/}) and ({\em b\/}) for the
instantaneous star formation rate per unit stellar mass ${\rm SFR}/M_{\star}$.
The models assumed a Salpeter IMF between 0.1 and $\rm 100~{\rm M_{\odot}}$,
solar metallicity, and the \citet{Cal00} extinction law.
\label{fig-resall2}
}
\end{figure}

Figure~\ref{fig-resall2} plots the variations of the derived extinctions
and instantaneous star formation rates per unit stellar mass with observed 
$J_{\rm s} - K_{\rm s}$ colours.  Results for both the standard CSF and
\tauint\ models are shown.  There are general trends of increasing $A_{V}$
and of decreasing ${\rm SFR} / M_{\star}$ with redder
$J_{\rm s} - K_{\rm s}$ colour.  The variations in derived ages (not
shown in Figure~\ref{fig-resall2}) are essentially the reverse of those
in the ${\rm SFR} / M_{\star}$ ratio, which provides a measure of
$1 / t_{\rm sf}$ (with a direct proportionality for the CSF model,
and a non-linear monotonic relation for the \tauint\ model).
These plots further show that if DRGs might be exceptional in their high
absolute SFRs, they are not at all in terms of their SFR per unit mass.

\subsection{Past History and Future Fate of DRGs}

\begin{figure}[p]
\epsscale{0.60}
\plotone{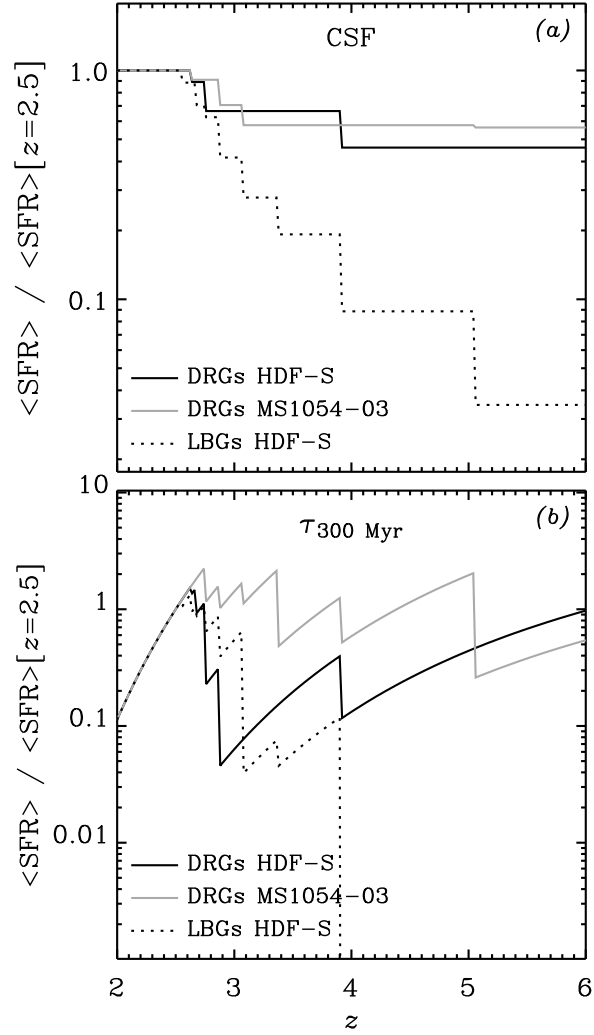}
\vspace{0.5cm}
\caption{
Relative evolution with redshift of the mean instantaneous star formation
rate for DRGs and LBGs at $2 \leq z \leq 3.5$ in the FIRES fields.
The different lines correspond to the DRG sample in HDF-S at
$K_{\rm s}^{\rm tot} < 22.5~{\rm mag}$ ({\em black solid line\/}),
the DRG sample in the $\rm MS\,1054-03$ field at
$K_{\rm s}^{\rm tot} < 21.7~{\rm mag}$ ({\em grey solid line\/}),
and the optically-selected LBG sample in HDF-S at
$K_{\rm s}^{\rm tot} < 22.5~{\rm mag}$ ({\em black dotted line\/}).
The curves were computed by averaging as a function of $z$ the scaled
model star formation rate $R(t)$ of the individual objects, assuming
they all lie at $z = 2.5$.  The curves are normalized to unity at
$z = 2.5$.  The models assumed a Salpeter IMF between 0.1 and
$\rm 100~M_{\odot}$, solar metallicity, and the \citet{Cal00}
extinction law.
({\em a\/}) Evolution from the results obtained assuming constant star
formation rate (CSF).
({\em b\/}) Evolution from the results obtained assuming the exponentially
declining $\tau_{\rm 300\,Myr}$ model.
\label{fig-cosmic_SFR}
}
\end{figure}

We here speculate about the evolution of the $2 \leq z \leq 3.5$ DRGs in
the FIRES fields, based on the results for the CSF and \tauint\ models.
Figure~\ref{fig-cosmic_SFR} plots the relative variations with redshift
of the average SFR of the DRG samples and, for comparison, of the HDF-S
LBG sample at $2 \leq z \leq 3.5$.
We computed the curves from the model SFR $R(t)$ used as input in the
evolutionary synthesis.  For each galaxy, we scaled the model $R(t)$
curves to match the derived instantaneous SFR at the epoch of observation.
For the purpose of this discussion, we in fact assumed that all galaxies
lie at $z = 2.5$, close to the mean and median redshift of both our DRG
and LBG samples (see \S~\ref{Sect-sample}).  We then normalized the
resulting average SFR curves to unity at $z = 2.5$.

For the CSF model, the average instantaneous SFR of the DRG samples has not
varied strongly since $z = 6$ up to $z = 2$, increasing by a factor of
$\approx 2$ with time.
On the other hand, the average instantaneous SFR of the HDF-S LBG sample has
increased by more than an order of magnitude over the same redshift interval.
For the \tauint\ model, the variations are much noisier but indicate, broadly
speaking, roughly comparable activity levels in the past out to $z \approx 4$,
where the mean instantaneous SFR of LBGs drops abruptly.
The relative variations in average SFR backwards in time are dominated by
the age distribution among each population, and the steeper drop for the
LBGs reflects their younger ages compared to the DRGs.
This extrapolation of the star formation histories suggests that
$2 \leq z \leq 3.5$ DRGs have started forming stars at very early epochs,
and earlier than $2 \leq z \leq 3.5$ LBGs have.  A possible scenario for
DRGs is that they are the evolved descendants of systems that experienced a
``Lyman-break phase'' at $z \ga 4$ \citep[see also][]{Fra03, Dok03, Dok04}.
This evolutionary picture may be consistent with the increased reddening in
rest-frame UV-to-optical colours inferred by \citet{Pap04} for Lyman-break
selected galaxies between $z \sim 4$ (``$B$-dropouts'') and $z \sim 3$
(``$U$-dropouts'').
We emphasize that the above discussion applies to the stellar
populations that dominate the rest-frame UV-optical emission
of the galaxies, as traced by the observed SEDs.

As discussed by \citet{Dok04}, the stellar masses of DRGs are already
very high at $z \sim 2 - 3$ and comparable to those of early-type galaxies
locally.  This may indicate that DRGs have already assembled most of their
stellar mass at these redshifts and evolve thereafter mostly passively.
Using our CSF model results, we estimated the stellar masses predicted at
$z = 0$ if star formation is maintained at the same rate as inferred at
the redshift of observation.  The median present-day stellar masses would
be $\rm 2 \times 10^{12}~M_{\odot}$, which correspond to the most massive
galaxies in local galaxy clusters.  While some of the DRGs might evolve
into such systems, it seems rather unlikely we have detected such large
numbers of their progenitors in our two small fields.
More plausibly, most of the DRGs probably become predominantly quiescent
systems at $z \la 2$, and may possibly grow further in mass during brief
merger events.  Assuming that DRGs stop forming stars around their redshift
of observation and evolve passively thereafter, the CSF models predict colours
of $I - K_{\rm s} > 4$ and $R - K_{\rm s} > 5$ at $1 \la z \la 2$, even if
all the dust is removed at the instant when star formation is quenched
(such red colours are also predicted using the \tauint\ models instead).
This may suggest an evolutionary connection between DRGs at $z > 2$ and
EROs at $z \la 1.5$.

In absolute terms, the instantaneous SFRs of $\rm \sim 100~M_{\odot}\,yr^{-1}$
derived from the SED modeling assuming CSF lie at the high end inferred for
the local galaxy population, in the regime of luminous and ultra-luminous
infrared galaxies.
Such a high SFR rate seems supported by the possible detection of one
of the non-AGN spectroscopically-confirmed DRG in the \mscl\ field at
sub-millimeter wavelengths from deep SCUBA observations
(\citealt{Dok04}; K. Kraiberg Knudsen \etal 2004, in preparation).
The CSF hypothesis is unverifiable with the data at hand, and may be
a valid representation for a few objects only.  Conversely, while the
\tauint\ model implies substantially lower instantaneous SFR at the
redshift of observation, the high stellar masses lead to extremely high
SFRs at earlier epochs ($\rm > 1000~M_{\odot}\,yr^{-1}$).
Obviously, the SFH of DRGs may well be more complex than our simplistic
assumptions.  The less model-dependent estimate from $M_{\star} / t_{\rm sf}$
of $\rm \sim 100~M_{\odot}\,yr^{-1}$ seems to suggest elevated levels
of star formation activity averaged over the lifetime of the stellar
populations traced by our SEDs.

\subsection{Relative Importance of DRGs and LBGs at $2 \leq z \leq 3.5$}
            \label{Sub-signif}

Using the results presented in this paper, we can crudely assess the relative
importance of the DRG and LBG populations at high redshift.  In the following,
we consider our samples of DRGs, LBGs, and $K_{\rm s}$-band selected sources
lying at $2 \leq z \leq 3.5$ and to the $K_{\rm s}$ total magnitude limits of
22.5 for HDF-S and 21.7 for the \mscl\ field.
In HDF-S, the LBGs outnumber the DRGs by a factor of three.
They contribute about 2.5 times more than DRGs to the total observed $K$-band
light and extrinsic rest-frame $V$-band luminosity $L_{V, \star}$ from all
sources, with fractions of $\approx 60\%$ and $\approx 25\%$, respectively.
The relative contributions are reversed for the stellar mass, where DRGs
account for $\approx 60\%$ of the total compared to $\approx 35\%$ for LBGs
(using the individual $M_{\star}$'s derived from either the CSF or \tauint\
models, and assuming the same model parameters for all populations).
In the \mscl\ field and to the corresponding brighter magnitude limit, the
DRGs produce about 35\% of the total observed $K_{\rm s}$-band light and
$L_{V, \star}$, and make up $\approx 65\%$ of the integrated $M_{\star}$.

This provides a first-order indication that DRGs may represent a significant
constituent of the $z \sim 2 - 3$ universe in terms of stellar mass.
Admittedly, our estimates are rough.  A detailed analysis of the
contribution of DRGs to the stellar mass budget in both FIRES fields will
be presented by G. Rudnick \etal (in preparation; see also \citealt{Rud03}
for first estimates in HDF-S).

\section{SUMMARY AND CONCLUSIONS}    \label{Sect-conclu}

We have presented the analysis and theoretical modeling of the
$\rm \lambda = 0.3 - 2.2~\mu m$ SEDs of 34 DRGs at $2 \leq z \leq 3.5$.
The galaxies were selected by their $J_{\rm s} - K_{\rm s} \geq 2.3$
colours in the HDF-S and \mscl\ fields observed as part of the FIRES
survey \citep{Fra00}, 
at $K_{\rm s} < 22.5$ and $\rm < 21.7~mag$, respectively.  The addition
of the \mscl\ field increases the area surveyed by a factor of five to an
$\rm \approx 0.7~mag$ brighter limit and triples the original sample from
HDF-S.  This allows for the first time a robust assessment of the
ensemble properties of DRGs.
We find large numbers of $J_{\rm s} - K_{\rm s}$ red objects in
both fields, with surface densities of $\rm \sim 1~arcmin^{-2}$ at
$K_{\rm s} < 21~{\rm mag}$.  
Since no $J_{\rm s} - K_{\rm s} \geq 2.3$ source is found in the HDF-N down
to $K_{\rm s} = 21~{\rm mag}$, cosmic variance is substantial.  Observations
of wider areas and multiple lines of sight will be required for a more
reliable determination of the surface density of the $J_{\rm s} - K_{\rm s}$
red population.

Using new diagnostic diagrams involving the $I_{814} - J_{\rm s}$,
$J_{\rm s} - H$, and $H - K_{\rm s}$ colours, we have shown that the red
NIR colours of the DRGs cannot be attributed to extinction effects alone and
require the presence of a prominent Balmer/4000\,\AA\ break characteristic of
evolved stars for a large fraction of them.  In the rest-frame optical, the
SEDs of DRGs fall within the envelope of normal nearby galaxies and, in the
rest-frame UV, indicate a wide range in star formation activity and/or dust
obscuration.  The rest-frame UV-to-optical SEDs of DRGs are in stark contrast
with those of LBGs, which are much bluer and more uniform.
These differences can be easily understood in terms of selection effects.
The LBG criteria preferentially select galaxies that are actively star-forming
and modestly obscured.  Our $J_{\rm s} - K_{\rm s} \geq 2.3$ criterion
primarily targets evolved systems, ranging from those with quiescent or no
star formation and little extinction to those that are highly obscured and
vigorously forming stars.

We have applied evolutionary synthesis models to the SEDs of our
DRGs to constrain quantitatively the age of the stellar populations, the
extinction, the stellar masses, and the star formation rates.  We have also
investigated the effects of variations in the assumed star formation history,
metallicity, and extinction law on the model results.  The derived properties
are similar for the HDF-S and \mscl\ field samples.
For constant star formation, a Salpeter IMF between 0.1 and 100\masssol, solar
metallicity, and the \citet{Cal00} extinction law, we obtain for each field a
median age $\rm \sim 2.0~Gyr$, extinction $A_{V} \sim 2.5~{\rm mag}$, stellar
mass $M_{\star} \sim 10^{11}~{\rm M_{\odot}}$, stellar mass-to-light ratio
$M_{\star}/L_{V,\,\star} \sim 1~{\rm M_{\odot}\,L_{V,\,\odot}^{-1}}$, and
instantaneous star formation rate ${\rm SFR} \sim 100~{\rm M_{\odot}\,yr^{-1}}$.
The stellar masses and M/L ratios are the most robust properties against
variations in the model parameters (neglecting those of the IMF) while the
instantaneous SFRs are strongly model-dependent.
Models with exponentially declining star formation rates and $e$-folding
timescales in the range $\rm \tau = 10~Myr - 1~Gyr$ generally lead to lower
ages, extinction values, and instantaneous SFRs but similar stellar masses
and M/L ratios within a factor of two.  The smallest median $A_{V}$'s are
derived with $\rm \tau = 300~Myr$.  For this \tauint\ model and the
assumptions above for the IMF, metallicity, and extinction law, the median
age and $A_{V}$ are 1~Gyr and 0.9~mag for both fields, and the median
SFRs $\rm \sim 10~{\rm M_{\odot}\,yr^{-1}}$.

An important implication of our work is that DRGs appear to be dominated
by significantly older, more obscured, and more massive stellar populations
than LBGs at similar redshifts and similar rest-frame $V$-band luminosities.
Using our sample of optically-selected LBGs in HDF-S, we have explicitely
shown that the distinction holds for constant star formation and \tauint\
models, the two star formation histories that bracket the range of median
$A_{V}$'s derived for our DRGs (neglecting variations in model parameters
between the two classes).  
For both DRGs and LBGs, there is a large spread in derived properties among
individual objects and some of the DRGs have ages, extinction, and stellar
masses characteristic of LBGs, and vice-versa.  However, the overlap is small,
especially for the stellar masses and M/L ratios.
Comparison with the LBG studies of \citet{Pap01}
and \citet{Sha01} is less straightforward because of differences in the
optical/NIR data sets, in the details of the sample selection, and in the
modeling but, broadly speaking, it supports the same conclusions.
More generally, we find that the derived age, extinction, and stellar
mass all increase with redder $J_{\rm s} - K_{\rm s}$ colour among the full
sample of $2 \leq z \leq 3.5$ $K_{\rm s}$-band-selected galaxies in the FIRES
fields, down to the same magnitude limits as the DRG samples.
The relationships are particularly well-defined for the stellar mass and
M/L ratio versus $J_{\rm s} - K_{\rm s}$ colour.

If DRGs represent the most evolved and most massive galaxies at $z \sim 2.5$,
it is tempting to speculate that they might be the precursors of the most
massive present-day galaxies.  Such a final fate has been proposed for LBGs
at $z \sim 3$ in the context of hierarchical models of galaxy formation
\citep[e.g.][]{Bau98}.  For DRGs, with stellar masses already similar to those
of local early-type galaxies, one may reason in a simplistic way that there
is little room for substantial growth through subsequent merger events.
Extrapolating the star formation histories backward in
time for both DRGs and LBGs at $2 \leq z \leq 3.5$, we infer that the DRG
population has been forming stars at a roughly constant rate and possibly
back to $z \sim 5 - 6$, whereas the LBGs started to form later at $z \sim 4$.
This is consistent with a scenario in which the DRGs are the evolved
descendants of systems undergoing a ``Lyman-break phase'' at $z > 4$.
The DRGs may evolve into the ERO population at $1 \la z \la 2$.
Another intriguing possibility is a connection with the high-redshift
population detected at sub-millimeter wavelengths.  One of the DRGs in the
\mscl\ field likely is the counterpart of a sub-millimeter source detected
with SCUBA (\citealt{Dok04}; K. Kraiberg Knudsen \etal 2004, in preparation).
In addition, although detailed information is currently available for a few
objects only, recent work indicates that some of the bright sub-millimeter
galaxies at $z = 2 - 3$ host evolved stellar populations, show evidence
of substantial chemical enrichment, and have dynamical/stellar masses
$\rm \sim 10^{11}~M_{\odot}$ \citep[e.g.][]{Gen03, Ner03, Tec04}.

The results presented here constitute a major improvement over our previous
studies of $J_{\rm s} - K_{\rm s}$ selected objects at $z > 2$.  Significant
uncertainties remain, however, and further investigations are clearly needed.
Our analysis still relies on a limited sample selected from two small fields
and the model parameters are largely unconstrained.  Deep photometric surveys
of wider areas and disjoint fields as well as extensive follow-up spectroscopy
will be crucial to assess (and average over) field-to-field variations, secure
the redshifts, derive robust number densities, and constrain more accurately
the physical properties (e.g., extinction, metallicity, dynamical masses). 
The current dearth of constraints on the nature of DRGs further limits our
ability to interpret their observed properties.  In that respect, it will be
very important to establish firmly the frequency of AGN among DRGs and their
contribution to the measured optical/NIR SEDs, in particular from spectroscopy.
It will also be important to obtain independent constraints on the current
and past star formation rates, with observations at infrared and submillimeter
wavelengths (e.g., with SCUBA and the Spitzer Space Telescope) complementing
rest-frame optical line diagnostics.
Ultimately, a better understanding of DRGs and their relationship to other
galaxy populations at high and low redshift may provide important constraints
on the assembly history of massive galaxies.

\acknowledgments
We would like to thank the ESO staff for their assistance and their efforts
in obtaining the high quality ISAAC and FORS1 data for FIRES and in making
them available to us.
We are grateful to Henk Hoekstra for kindly computing for us the lensing
magnifications in the \mscl\ field.
We thank the referee, Casey Papovich, for very useful comments and suggestions
that helped improve the paper.
N.M.F.S. appreciates the critical discussions and encouraging comments
of Matt Lehnert.
N.M.F.S. acknowledges the Department of Astronomy at Yale University for
its hospitality and generous support during a working visit, and the
Leids Kerkhoven-Bosscha Fonds for travel support.



%\appendix

\clearpage

%% Here figures for preprint version.

%\setcounter{figure}{0}

\clearpage

%\clearpage

\clearpage

\clearpage

\clearpage

\clearpage

\clearpage

\clearpage

\clearpage

\clearpage

\clearpage

\clearpage

\clearpage

\clearpage

\begin{deluxetable}{lcccc}
\tabletypesize{\small}
\tablewidth{250pt}
\tablenum{1}
\tablecolumns{5}
\tablecaption{Summary of selected properties of the adopted
        $J_{\rm s} - K_{\rm s} \geq 2.3$ samples at $2 \leq z \leq 3.5$
        in the HDF-S and $\rm MS\,1054-03$ fields
        \label{tab-sample}}
\tablehead{
\multicolumn{5}{c}{HDF-S: 11 DRGs at $K_{\rm s}^{\rm tot} < 22.5~{\rm mag}$}
}
\startdata
\multicolumn{2}{c}{Property} & Median & Mean & rms  \\
\cline{1-5} \\
$K_{\rm s}^{\rm tot}$    & (mag)  &  21.39  &  21.25  &  0.55  \\
$V^{\rm tot}$            & (mag)  &  26.33  &  26.69  &  0.96  \\
$J_{\rm s} - K_{\rm s}$  & (mag)  &   2.60  &   2.75  &  0.43  \\
$z$                      & \ldots &   2.50  &   2.60  &  0.42  \\
\cline{1-5} \\
\multicolumn{5}{c}{$\rm MS\,1054-03$: 
                   23 DRGs at $K_{\rm s}^{\rm tot} \leq 21.7~{\rm mag}$} \\
\cline{1-5} \\
\multicolumn{2}{c}{Property} & Median & Mean & rms  \\
\cline{1-5} \\
$K_{\rm s}^{\rm tot}$    & (mag)  &  20.95  &  20.84  &  0.74  \\
$V^{\rm tot}$            & (mag)  &  26.86  &  26.61  &  0.94  \\
$J_{\rm s} - K_{\rm s}$  & (mag)  &   2.79  &   2.85  &  0.43  \\
$z$                      & \ldots &   2.42  &   2.44  &  0.30  \\
\enddata
\tablecomments{
All magnitudes are expressed in the Vega system.
The $K_{\rm s}^{\rm tot}$ and $V^{\rm tot}$ data for the $\rm MS\,1054-03$
field sample are corrected for the (small) lensing magnifications.
}
\end{deluxetable}

%\clearpage

\begin{deluxetable}{llcccccc}
\tabletypesize{\small}
\tablecolumns{8}
\tablewidth{450pt}
\tablenum{2}
\tablecaption{Summary of model results for $J_{\rm s} - K_{\rm s} \geq 2.3$
        samples in the HDF-S and $\rm MS\,1054-03$ fields\,\tablenotemark{a}
        \label{tab-res_std2}}
\tablehead{
  \multicolumn{8}{c}{HDF-S}
}
\startdata
  & & \multicolumn{3}{c}{CSF} & \multicolumn{3}{c}{$\tau_{\rm 300\,Myr}$} \\
\multicolumn{2}{c}{Property} & Median & Mean & rms & Median & Mean & rms \\
  \cline{1-5} \cline{6-8} \\
Age $t_{\rm sf}$        & (Gyr)  &
   1.7  & 1.5  & 0.86  & 1.0  & 1.1  & 1.0  \\
SFR-weighted age $\langle t \rangle_{\rm SFR}$        & (Gyr)  &
   0.85 & 0.74 & 0.43  & 0.75 & 0.93 & 0.92 \\
$A_{V}$                & (mag)  &
   2.7  & 2.0  & 0.78  & 0.90 & 1.20 & 0.96 \\
$M_{\star}$             & ($\rm 10^{10}~M_{\odot}$)  &
   8.1  & 15   & 13    & 7.5  & 7.9  & 5.7  \\
$L_{V,\star}$\,\tablenotemark{b}       & ($10^{10}~{\rm L}_{V,\odot}$)  &
   5.0  & 5.7  & 3.6   & 5.0  & 5.7  & 3.6  \\
$M_{\star}/L_{V,\star}$ & (${\rm M_{\odot}}\,{\rm L}_{V,\odot}^{-1}$)  &
   1.2  & 2.7  & 2.2   & 1.2  & 1.5  & 0.89 \\
$M_{\star}/t_{\rm sf}$  & ($\rm M_{\odot}\,yr^{-1}$)  &
   91   & 120  &  96   & 91   & 150  & 160  \\
$\rm SFR$               & ($\rm M_{\odot}\,yr^{-1}$)  &
  120   & 150  & 120   & 6.7  & 96   & 140  \\
${\rm SFR}/M_{\star}$ & ($\rm 10^{-10}~yr^{-1}$) &
  7.6   & 19   & 26    & 1.5  & 18   & 25   \\
${\rm SFR}/L_{V,\star}$ & 
($\rm 10^{-10}~M_{\odot}\,yr^{-1}\,{\rm L}_{V,\odot}^{-1}$) &
  30    & 33   & 30    & 1.2  & 21   & 32   \\
\cline{1-8} \\ 
\multicolumn{8}{c}{$\rm MS\,1054-03$} \\
\cline{1-8} \\ 
  & & \multicolumn{3}{c}{CSF} & \multicolumn{3}{c}{$\tau_{\rm 300\,Myr}$} \\
\multicolumn{2}{c}{Property} & Median & Mean & rms & Median & Mean & rms \\
  \cline{1-5} \cline{6-8} \\
Age $t_{\rm sf}$        & (Gyr)  &
   2.0  & 1.7  & 0.84  & 1.0  & 1.3  & 0.91 \\
SFR-weighted age $\langle t \rangle_{\rm SFR}$        & (Gyr)  &
   1.0  & 0.83 & 0.42  & 0.75 & 1.1  & 0.85 \\
$A_{V}$              & (mag)  &
   2.4  & 2.2  & 0.56  & 0.90 & 1.3  & 0.80 \\
$M_{\star}$             & ($\rm 10^{10}~M_{\odot}$)  &
   16   & 21   & 16    & 10   & 15   & 11   \\
$L_{V,\star}$\,\tablenotemark{b}       & ($10^{10}~{\rm L}_{V,\odot}$)  &
   5.1  & 7.4  & 6.4   & 5.1  & 7.4  & 6.4  \\
$M_{\star}/L_{V,\star}$ & (${\rm M_{\odot}}\,{\rm L}_{V,\odot}^{-1}$)  &
   2.3  & 3.3  & 1.9   & 1.8  & 2.4  & 1.7  \\
$M_{\star}/t_{\rm sf}$  & ($\rm M_{\odot}\,yr^{-1}$)  &
  140   & 150  & 110   & 94  & 170  & 150  \\
$\rm SFR$               & ($\rm M_{\odot}\,yr^{-1}$)  &
  170   & 190  & 130   & 23   & 69   & 110  \\
${\rm SFR}/M_{\star}$ & ($\rm 10^{-10}~yr^{-1}$) &
  6.5   & 15   & 19    & 1.5  & 6.9  & 10   \\
${\rm SFR}/L_{V,\star}$ & 
($\rm 10^{-10}~M_{\odot}\,yr^{-1}\,{\rm L}_{V,\odot}^{-1}$) &
  31    & 32   & 22    & 2.7  & 11   & 17   \\
\enddata
\tablenotetext{a}
{
The samples selected with $2 \leq z \leq 3.5$, with
$K_{\rm s} < 22.5~{\rm mag}$ for HDF-S and
$K_{\rm s} < 21.7~{\rm mag}$ for the $\rm MS\,1054-03$ field.
The main model parameters are a Salpeter IMF between 0.1 and 
$100~{\rm M_{\odot}}$, solar metallicity, the extinction law
of \citet{Cal00}, and either a star formation rate constant
in time (CSF) or exponentially declining in time with $e$-folding
timescale $\rm \tau = 300~Myr$ ($\tau_{\rm 300\,Myr}$).
}
\tablenotetext{b}
{
The $L_{V,\star}$ values were obtained independently, based on the
less model-dependent method described by \citet{Rud03}, and are the
dust-attenuated luminosities.  These estimates are used in all ratios
involving $L_{V,\star}$.  The ages, extinctions, stellar masses, and 
instantaneous SFRs were derived from the best-fit model to the SEDs.
}
\end{deluxetable}

\end{document}